\documentclass[aps,pre,preprint,groupedaddress,amsmath,amssymb]{revtex4-1}

\usepackage{graphicx}
\usepackage{dcolumn}
\usepackage{bm}
\usepackage{xcolor}
\definecolor{greenRev1}{rgb}{0.1,0.6,0.1}
\definecolor{blueRev1}{rgb}{0.1,0.1,0.6}

\begin{document}


\title{Analysis of Lundgren's matched asymptotic expansion approach to the K\'arm\'an-Howarth equation using the EDQNM turbulence closure}

\author{M. Meldi}
\email{Corresponding Author: marcello.meldi@ensma.fr}
\affiliation{Institut PPRIME, Department of Fluid Flow, Heat Transfer and Combustion, ENSMA - CNRS - Universit\'{e} de Poitiers, UPR 3346, SP2MI - T\'{e}l\'{e}port, 211 Bd. Marie et Pierre Curie, B.P. 30179 F86962 Futuroscope Chasseneuil Cedex, France}
\author{J.~C. Vassilicos}%
\affiliation{Univ. Lille, CNRS, ONERA, Arts et M\'{e}tiers ParisTech, Centrale Lille, UMR 9014 - LMFL - Laboratoire de M\'{e}canique des fluides de Lille - Kamp\'{e} de Feriet, F-59000 Lille, France
}%

\date{\today}

\begin{abstract}

In this paper we investigate whether the features of the
non-equilibrium cascade, which have been identified in recent studies
using high-fidelity tools, can be captured in the case of the
classical dissipation scaling by turbulence closures based on the
statistical description of freely decaying isotropic
turbulence. Numerical results obtained using the EDQNM model over a
very large range of Reynolds numbers (from $Re_{\lambda}=50$ up to
$Re_{\lambda}=10^6$) are analyzed to perform an extensive
investigation of the scaling region identified as inertial range in
Kolmogorov's theory. It is observed that EDQNM results are in
agreement with the results of Lundgren's matched asymptotic expansion
approach to the K\'arm\'an-Howarth equation. Both predict that the
Kolmogorov inertial range equilibrium {is never
  obtained} irrespective of Reynolds number. Equilibrium is
reached in the vicinity of the Taylor
length $\lambda$ (which depends on viscosity) as Reynolds number tends
to infinity and there is a gradual departure from equilibrium as the
length scale moves away from $\lambda$, in particular towards scales
larger than $\lambda$ all the way to the integral
length-scale.
 
\end{abstract}

\maketitle


\section{Introduction}

A new turbulence dissipation scaling has been discovered over the past
10 years (\cite{Vassilicos2015_arfm,Cafiero2019_prsa} and references
therein) which characterises the non-equilibrium cascade in the
relative near-field of various turbulent flows: grid-generated
turbulence, turbulent jets, turbulent wakes and periodic turbulence
(both forced and decaying). This relative near-field is in fact quite
extensive in practical terms and the turbulence dissipation scaling in
that relative near-field incorporates an explicit dependence on
inlet/initial conditions. {This scaling can be
  written as $\varepsilon \sim U_0 L_0 \mathcal{K} / L^2$ where
  $\varepsilon$ is the energy dissipation rate, $\mathcal{K}$ is the
  turbulent kinetic energy and $L$ is the integral length scale. On
  the other hand, $U_0$ and $L_0$ are a velocity scale and a length
  scale associated with initial/inlet conditions.} Further downstream
one expects no such explicit dependence, and a turbulence dissipation
scaling which is the classical Taylor-Kolmogorov scaling
{$\varepsilon \sim \mathcal{K}^{3/2} / L$.}
\cite{Taylor1935_prsl,Kolmogorov1941_dan,Antonia2006,Vassilicos2015_arfm}.
Such a transition from one dissipation scaling to another has indeed
been observed in some turbulent flows
\cite{Vassilicos2015_arfm,Goto2016_pre}. However, the recent work of
Goto \& Vassilicos \cite{Goto2016_pre} suggests that the far-field
dissipation scaling is not necessarily the result of an equilibrium
turbulence cascade even if it takes the exact same form as it would
take in an equilibrium cascade situation.

Obligado \& Vassilicos \cite{Obligado2019_epl} explained why there is
no equilibrium cascade at length-scales large enough to justify a
Taylor-Kolmogorov dissipation scaling in the case of freely decaying
homogeneous isotropic turbulence (HIT) with classical
Taylor-Kolmogorov dissipation scaling. {The
  starting point is Von Karman \& Howarth \cite{Karman1938_prsA}
  equation in the form that it takes when expressed for structure
  functions as in Landau \& Lifschitz \cite{Landau1987}:}
 \begin{equation}
  -{2\over 3} \varepsilon -{1\over 2} {\partial \over \partial t} S_2
  = {1\over 6r^{4}}{\partial r^{4} S_{3}\over \partial r} -{\nu\over
    r^{4}}{\partial \over \partial r}(r^{4}{\partial S_{2} \over
    \partial r})
   \label{KH}
\end{equation}
{where $\varepsilon$ is the turbulence
  dissipation rate, $\nu$ is the kinematic viscosity of the fluid, $r$
  is the distance between the pairs of points defining the second and
  third order longitudinal structure functions $S_2$ and $S_3$
  respectively.}
Using Lundgren's \cite{lundgren2002} analysis of the
K\'arm\'an-Howarth equation \ref{KH} based on matched
asymptotic expansions and wind tunnel data for grid-generated
approximate HIT conditions, they demonstrated that spectral
equilibrium can only be achieved in the vicinity of the Taylor
length-scale when the Reynolds number tends to infinity. For any
Reynolds number, including in the limit of infinite Reynolds number,
the cascade remains out of equilibrium at any other length-scale which
is not a fixed multiple of $\lambda$. The larger the length-scale
compared to $\lambda$ the wider the departure from
equilibrium. Without equilibrium at length-scales large enough to be
independent of viscosity it is not possible to justify the
Taylor-Kolmogorov dissipation scaling by a balance between the
turbulence dissipation $\varepsilon$ and the large-scale interscale
energy flux (or rate of energy injection into the cascade at the large
scales). Goto \& Vassilicos \cite{Goto2016_pre} justified the
far-field Taylor-Kolmogorov dissipation scaling on the basis of their
concept of balanced non-equilibrium to which we return at the end of
this paper.

The features of the non-equilibrium cascade have been up to now
studied using high-fidelity tools such as experiments and direct
numerical simulations.
In this paper we investigate whether features of the non-equilibrium
cascade can be captured by turbulence closures based on the
statistical description of freely decaying isotropic turbulence. To
this purpose, the EDQNM model \cite{Orszag1970_jfm} has been
identified. This tool can be used to investigate the statistical
features of homogeneous turbulence and in particular the time
evolution of freely decaying isotropic turbulence in a far field state.
\cite{Bos2007_pof,Lesieur2008_springer,Meldi2013_jot,Sagaut2018_springer}. One
of the main favorable features of this model is that it can be used to
investigate freely decaying HIT at Reynolds numbers which are not
reachable by high-fidelity tools. The present analysis provides
answers to two main questions. The first one is whether the EDQNM is
consistent with the absence of interscale equilibrium at all Reynolds
numbers in the far field except at scales in the vicinity of the
Taylor length scale. The second one is if there is agreement between
EDQNM and Lundgren's \cite{lundgren2002} matched asymptotic expansion
analysis of the Karman-Howarth equation.

The paper is organized as follows. In section 2 we describe the EDQNM
model we use and in section 3 we present our results and the answers
to the questions we posed in this introduction. We conclude in section
4.

\section{The EDQNM model}
\label{sec:EDQNM}  

The EDQNM model
\citep{Orszag1970_jfm,Lesieur2008_springer,Sagaut2018_springer} is a
turbulence closure in spectral space. It resolves the numerical discretization of the classical Lin
equation:
\begin{equation}
\label{eq:Lin_F_freeDec}
    \frac{\partial E(k,t)}{\partial t} + 2 \nu k^2 E(k,t) = T(k,t) + F(k,t)
\end{equation}
{where $F(k,t)$ is a forcing term in the spectral
  space to be explicitely provided. Present calculations are performed
  for freely decaying HIT, so $F(k,t)=0$.}  $T(k,t)$ is the non-linear energy
transfer. This term is related to the interscale energy flux via the
expression $\Pi(k,t)=\int_k^{+\infty} T(k,t) dk$. Integrating the Lin
equation one obtains an equation where $\Pi(k,t)$ appears rather than
$T(k,t)$ and which is the spectral space equivalent of the
K\'arm\'an-Howarth equation \ref{KH} (see
\cite{lundgren2002} and references therein). The EDQNM model is based
on two fundamental assumptions applied to the evolution equation of
the three-point third-order velocity correlation, which is used to
calculate $T(k,t)$:
\begin{enumerate}
\item The quasi-normal (QN) approximation is used to express
  fourth-order moments by a sum of products of second-order
  moments. The fourth-order cumulants (i.e. deviation from a Gaussian
  pdf of the velocity derivatives) are represented via a linear
  damping term which is governed by the eddy damping rate $\eta_E$.
\item A Markovianization procedure is used assuming that the relaxation time of the third-order correlations is small when compared with the relaxation time of the second-order correlations.
\end{enumerate}

Using these two hypotheses, a closed expression for $T(k,t)$ is
obtained:

\begin{equation}
T(k,t)= \int_{p+q=k} \Theta_{kpq} (xy+z^3) E(q,t) [E(p,t)pk^2 -E(k,t)p^3] \frac{dp dq}{pq}
\label{eq:T_EDQNM}
\end{equation}

where $[k,p,q]$ represents the spectral system of coordinates and
$xy+z^3$ is a geometric factor determined by the shape of the
triangles respecting the condition $p+q=k$ in the spectral space. The
term $\Theta_{kpq}^{-1} = \eta_{E} (k,t) + \eta_{E} (p,t) + \eta_{E}
(q,t)$ is a spectral time scale which is inversely proportional to the
eddy damping rate $\eta_E$. This term is usually derived by the
following model relation \cite{Pouquet1975_jfm}:
\begin{equation}
\label{eq:eta_E}
\eta_E(k,t) = A \, \sqrt{\int_0^k \, p^2 E(p,t) \, dp} + \nu k^2.
\end{equation}

The free coefficient $A \in [0.3, \, 0.5]$ is usually set to optimize the value for the constant $C_K$ governing $E$ in the scaling region i.e. $E(k,t)= C_K \varepsilon^{2/3} k^{-5/3}$ \cite{Andre1977_jfm}. A different model for the determination of the eddy damping term is the EDQNM-LMFA proposed by Bos \& Bertoglio
\cite{Bos2006_pof}. In this model, the eddy damping
rate $\eta_E$ is calculated using a model equation for the velocity
derivative cross-correlation $\mathcal{F}_{CC}$, so that
$\eta_{E}(k,t)=E(k,t)/\mathcal{F}_{CC}(k,t)+ \nu k^2$.



The turbulent kinetic energy $\mathcal{K}(t)$, the energy dissipation
rate $\varepsilon(t)$ and all the other physical quantities are
derived via manipulation or integration of $E(k,t)$ and $T(k,t)$. The
calculations are performed using an adaptive spectral mesh strategy
\citep{Meldi2014_jcp} which preserves both the large-scale and the
small-scale resolution. {This operation is
  performed by updating the mesh elements so that the conditions
  $k_L(t) / k_{min}(t) = s_{L} = const$ and $k_{\eta}(t) / k_{max}(t)
  = s_{\eta} = const$ are verified. Here,} $k_L(t)= L^{-1}(t)$ is the
wave number associated with the integral length scale $L$, $k_\eta(t)=
\eta^{-1}(t)$ is the wave number associated with the Kolmogorov scale
$\eta$, $k_{max}$ is the largest resolved mode and $k_{min}$ is the
smallest resolved mode. This strategy allows for the complete control
of confinement effects. {For every calculation
  $s_{\eta}=10^{-1}$ is imposed, which implies that the smallest
  resolved scale has a size of $\approx 0.1 \eta$. More details about
  the value of the constant $s_{L}$ are provided below.}

{A database of EDQNM calculations has been
  established in order to quantify the sensitivity of the physical
  quantities investigated to variations in the set up of the problem
  as detailed in the following list.}

\begin{itemize}
    \item{ {\underline{Initial
          conditions}. Different proposals have been employed to
        initialize the energy spectrum at time $t=0$ of the free decay
        simulations. The first one is inspired from the functional
        form found in Pope \cite{Pope2000_cambridge} and Meyers \&
        Meneveau \cite{Meyers2008_pof}:}
\begin{equation}
E_I(k,t=0)= C_K \, {\varepsilon}^{2/3} k^{-5/3} f_L (k L) f_{\eta} (k \eta)
\label{eq:EnergySp_Pope}
\end{equation}
{with}
\begin{equation}
\label{eq:largescale_Pope}
f_L (k L) = \left( \frac{k L}{[(k L)^{1.5} + c_L]^{1/1.5}} \right)^{5/3 + \sigma} \: , \hspace{0.05cm} f_{\eta}(k \eta) = \exp(-\beta([(k\eta)^4+ c_{\eta}^4]^{1/4} -c_{\eta})) 
\end{equation}
{where $C_K \approx 1.6$ is the Kolmogorov
  constant. The dimensionless coefficients in equations
  \ref{eq:EnergySp_Pope} - \ref{eq:largescale_Pope} have been set to
  $c_{\eta}=0.4$, $\beta=5.3$; $c_L$ has been chosen such as to obtain
  $L(0)=1$. The parameter $\sigma$, which controls the shape of the
  energy spectrum at the large scales, is discussed separately
  analyzed in the next bullet point. A second functional form employed
  to initialise the energy spectrum at $t=0$ is the classical
  exponential relation
\begin{equation}
E(k,t=0)= k^\sigma \exp((-\sigma/2)k^2)
\label{eq:EnergySp_Exp}
\end{equation}
which is controled by the parameter $\sigma$ only. }

  
}
\item{{\underline{Large scale parameter
      $\sigma$}. This parameter is strictly concerned with the
    initialization procedure described above, but the effects of
    $\sigma$ on HIT decay do not disappear if sufficient resolution is
    provided at the large scales
    \cite{Meldi2012_jfm,Meldi2017_jfm}. The values chosen for this
    parameter are $\sigma=2,4$ which correspond to the well known
    cases of Saffman turbulence and Loitsiansky turbulence,
    respectively.}}
\item{{\underline{EDQNM model}. Runs have been
    perfomed using both the classical version of EDQNM and the
    EDQNM-LMFA model.}}
\item{{\underline{Resolution at the large
      scales}. The last aspect of the sensitivity analysis deals with
    confinement i.e. lack of resolution at the large scales. Most of
    the simulations in the database are run using $s_{L}=500$, which
    ensures that confinement effects are completely excluded. One
    simulation is instead run for $s_{L}=10$, in order to investigate
    the effects of saturation over the physical quantities
    investigated.}}
\end{itemize}

\begin{table}[ht]
\centering
\begin{tabular*}{\textwidth}{c @{\extracolsep{\fill}} lcccc}
\hline
Simulation Nr. & Functional form ($t=0$) & $\sigma$ & EDQNM model & $k_L(t)/k_{min}(t)$ \\
\hline
$1$ & Pope (eq. \ref{eq:EnergySp_Pope} - \ref{eq:largescale_Pope}) & $2$ & LMFA & $500$ \\
$2$ & Pope & $2$ & LMFA & $10$ \\
$3$ & Pope & $4$ & LMFA & $500$ \\
$4$ & Pope & $4$ & classical & $500$ \\
$5$ & exponential (eq. \ref{eq:EnergySp_Exp}) & $2$ & classical & $500$ \\
$6$ & exponential & $4$ & classical & $500$ \\
$7$ & exponential & $4$ & LMFA & $500$ 
\end{tabular*}
\caption{\label{table::1} {Database of EDQNM calculations used in the present analysis.}}
\end{table}

{A summary of the features of the calculations
  included in the database is reported in table
  \ref{table::1}. }{Every calculation of the database }is performed using
an initial Reynolds number of $Re_{\lambda}(t=0)= 10^6$. A transient
regime is initially observed, which is governed by the features of the
functional form prescribed for $t=0$. During this transient, the
Reynolds number increases up to $Re_{\lambda} \approx 2 \times 10^6$
at $t \approx t_0$, where $t_0=\mathcal{K}(t=0)/ \varepsilon(t=0)$ is
the initial turnover time. After this first phase, the statistics
progressively lose memory { of those initial
  condition prescribed in the scaling range and in the small scale
  region.} A classical power law decay is then observed
\citep{Comte-Bellot1966_jfm,Meldi2012_jfm},
{which is governed by the parameter $\sigma$
  prescribed in the functional form of $E$. This parameter controls
  the slope of the energy spectrum at the very large scales}. Data in
the form of $E(k,t)$ are sampled in the range $t/t_0 \in [10^3, \,
  10^{C}] $, which corresponds to a free decay from
$Re_{\lambda}(t/t_0 \approx 10^{3}) = 10^6$ to $Re_{\lambda}(t/t_0
\approx 10^C) = 50, \, C \in[25,45]$. {The time
  range chosen for sampling varies with initial conditions and with
  value of $\sigma$ in particular.}

\begin{figure}[h!]
\begin{tabular}{cc}
\includegraphics[width=0.48\linewidth]{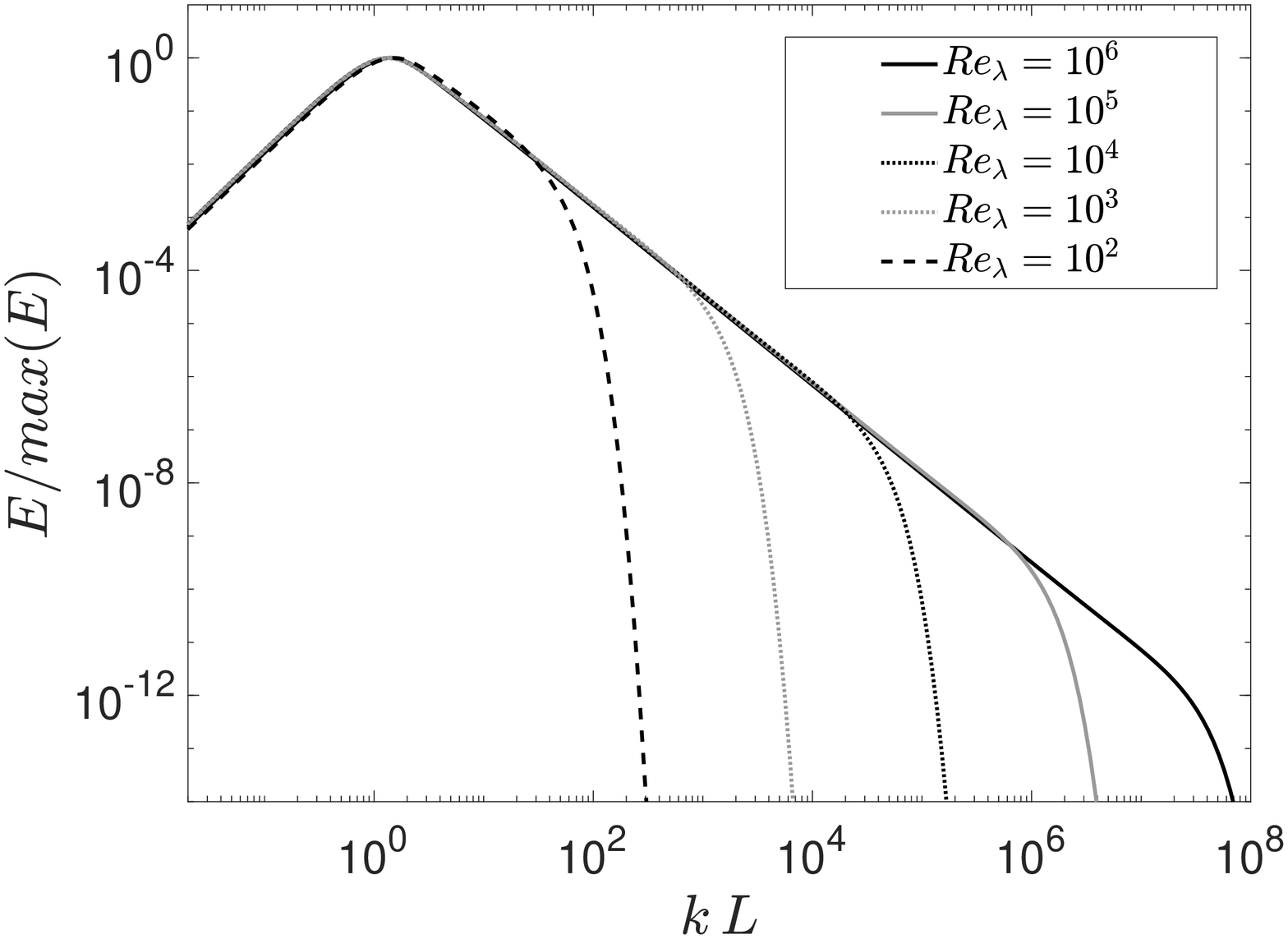} & \includegraphics[width=0.48\linewidth]{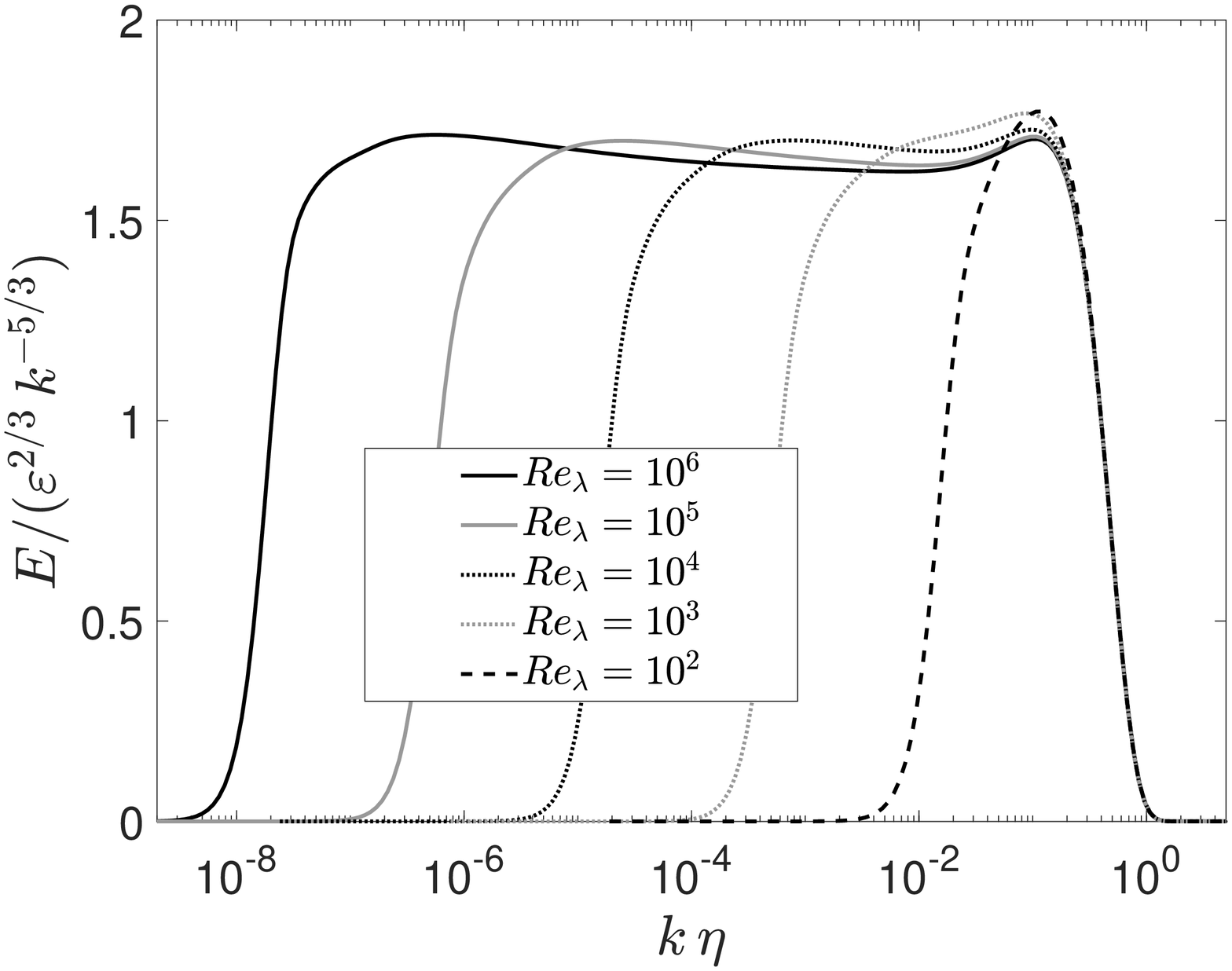} \\
(a) & (b) \\
\includegraphics[width=0.48\linewidth]{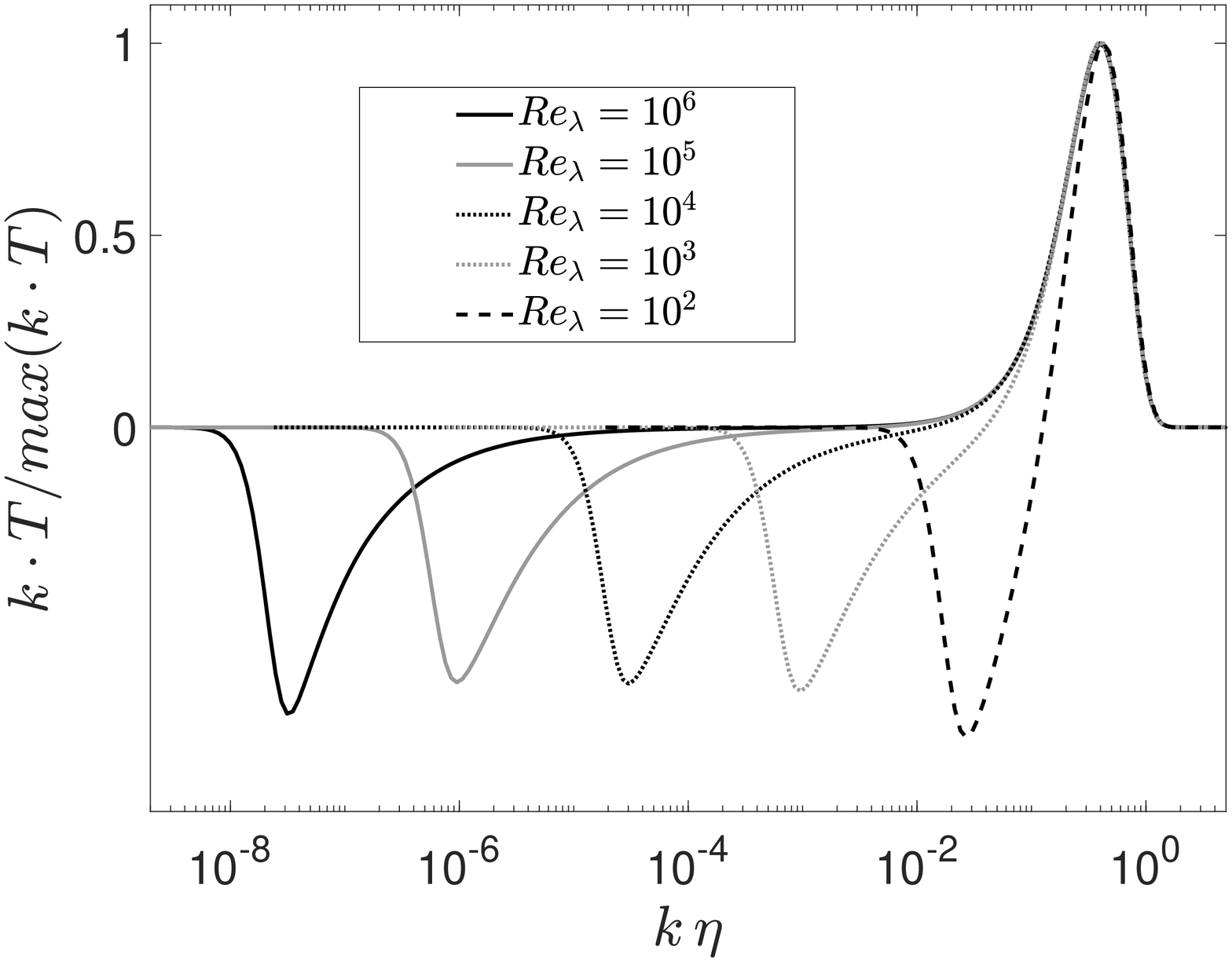} & \includegraphics[width=0.48\linewidth]{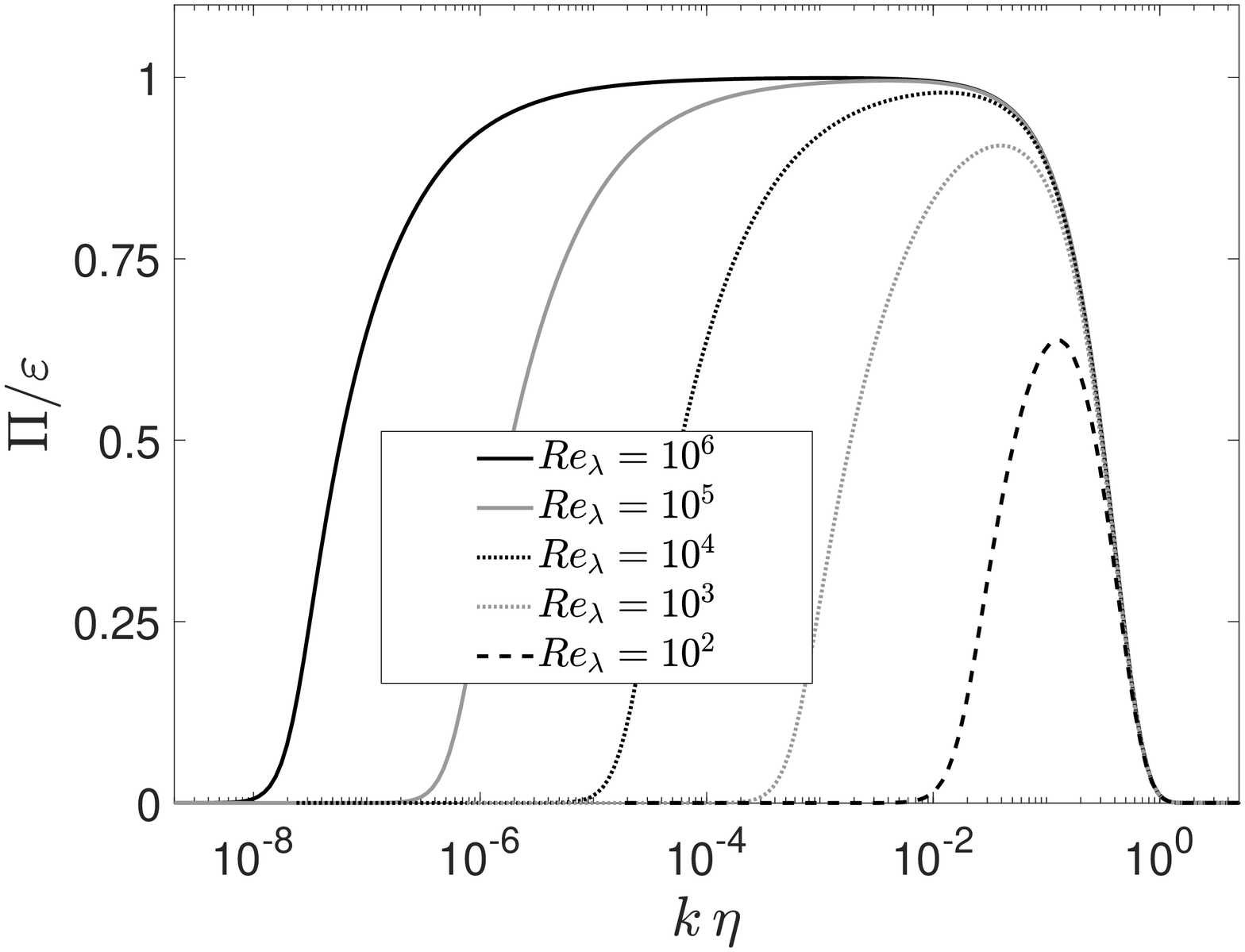} \\
(c) & (d)
\end{tabular}
\caption{\label{fig:1} Energy spectrum and non-linear energy transfer
  calculated via EDQNM. (a) Normalized energy spectrum $E/max(E)$. (b)
  Compensated energy spectrum $E/(k^{-5/3} \varepsilon^{2/3}$). (c)
  Normalized derivative in Fourier space of the non-linear energy
  transfer budget term, $k T/max(k T$). (d) Normalized inter scale
  energy flux $\Pi / \varepsilon$.}
\end{figure}

{Results are now discussed for simulation 1 of
  the database. For this simulation the effects of the functional form
  prescribed at $t=0$, which are measured via the rate of convergence
  towards the expected power-law decay exponent, can be estimated to
  be $\approx 5\%$ for $Re_{\lambda} = 10^6$ and $<0.1\%$ for
  $Re_{\lambda} = 10^5$.} The energy spectrum $E$, the compensated
energy spectrum $E / (\varepsilon^{2/3} k^{-5/3})$, the derivative of
the non-linear energy transfer $T$ and the interscale energy flux
$\Pi$ are shown for $Re_{\lambda}=10^6, \, 10^5, \, 10^4, \, 10^3, \,
10^2$ in figure \ref{fig:1}. The EDQNM results we obtain comply with
previous EDQNM results \cite{Bos2007_pof,Meldi2013_jot} and, plotted
as in this figure and without further analysis, may appear to support
the picture drawn by the $K41$ theory
\cite{Kolmogorov41a,Kolmogorov41b} for $Re_{\lambda} \to +\infty$. The
compensated energy spectrum as well as the non-linear energy transfer
collapse remarkably well in the small scale region when plotted
against $k \eta$. In addition, one can see the emergence of a scaling
range for sufficiently high $Re_{\lambda}$ (larger than $10^4$
according to present data). This range, which can be described by the
relation $E(k) = C_K \varepsilon^{2/3} k^{-5/3} \approx 1.6
\varepsilon^{2/3} k^{-5/3}$ to a close approximation, appears to be
characterized by a close-to-zero derivative of the non linear energy
transfer and by $\Pi / \varepsilon \approx 1$, which one may infer
from figure \ref{fig:1}(d).

The EDQNM predictions for $E$ and $T$ are used to obtain expressions
for second-order and third-order structure functions. These quantities
can be calculated exactly via the following integral / differential
relations \citep{Tchoufag2012_pof,Bos2012_pof}:
\begin{eqnarray}
\label{eq:S2_EDQNM}
S_2(r,t) &=& \int_0^{+\infty} 4 \, E(k,t) \, \left[\frac{1}{3} - \frac{sin(kr) - (kr) cos(kr)}{(kr)^3} \right] \, dk \\
\label{eq:S3_EDQNM}
S_3(r,t) &=& \int_0^{+\infty} 12 \, T(k,t) \, \frac{3 (sin(kr) - (kr) cos(kr))- (kr)^2 sin(kr)}{(kr)^5}  \, dk \\
\label{eq:f_nonEq}
f(r,t)&=& - \frac{1}{\varepsilon} \frac{\partial S_2(r,t)}{\partial t} \\
\label{eq:F_nonEq}
F(r,t)&=& -  \frac{3}{\varepsilon r^5} \int_0^r  {r^{\prime}}^4 \frac{\partial S_2}{\partial t} \, dr^{\prime}
\end{eqnarray}

where $S_2$ is the second-order moment of the longitudinal velocity
increment, $S_3$ is the third-order moment and $f$ and $F$ are
non-stationarity functions \cite{Obligado2019_epl}. These
non-stationarity functions measure the departure from equilibrium
scale by scale and appear naturally from the form of the
K\'arm\'an-Howarth equation (\ref{KH}) given by Landau \& Lifschitz
\cite{Landau1987} (see also \cite{Danaila99,lundgren2002}). The $-4/5$
law $S_3 \approx -{4\over 5}\varepsilon r$ might be derived from this
equation in a range of scales $r$ where $F\approx 0$ (implying
equilibrium) and viscous diffusion is negligible. The function $F$ is
obtained by a normalised integration of the non-stationarity function
$f$ which is directly interpretable in terms of equilibrium as it is
small at a given scale $r$ and a given time $t$ if $\frac{\partial
  S_2(r,t)}{\partial t}$ is small compared to ${\varepsilon}$. The
question which was raised by
Obligado \& Vassilicos \cite{Obligado2019_epl} concerns the range of
scales where one might safely assume $F\approx 0$ and this is part of
the questions which we now address in terms of EDQNM.
 
\section{Results}
\label{sec:results}

\begin{figure}[h!]
\includegraphics[width=0.8\linewidth]{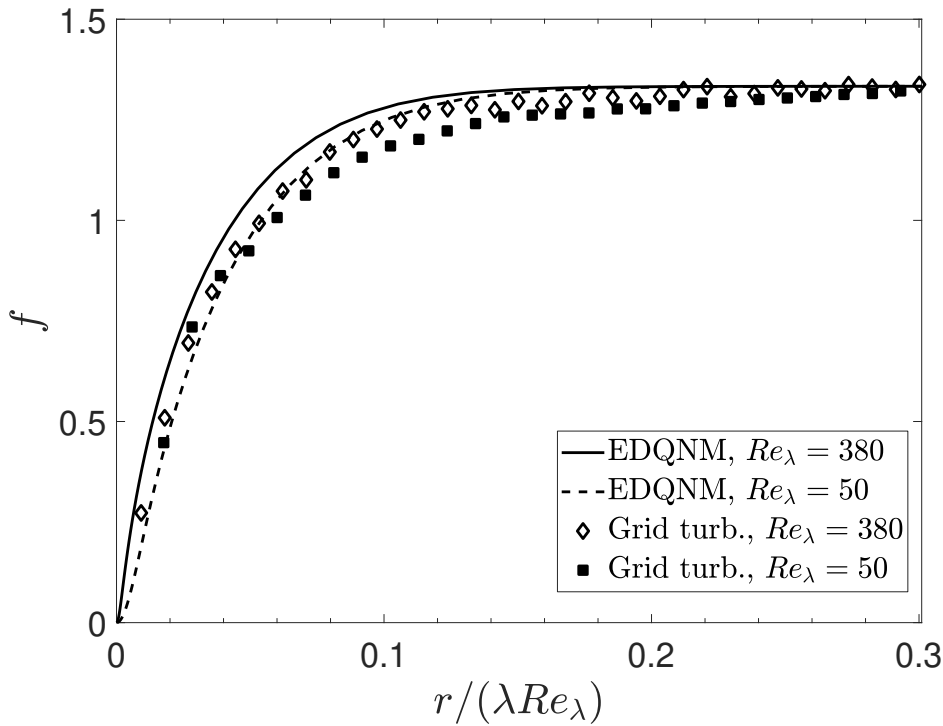}
\caption{\label{fig:1b} Non-stationarity function $f$. EDQNM results
  are compared with experimental grid turbulence results.}
\end{figure}

Results {presented in this section are again taken from simulation 1 of the database. Data} are first compared with
experimental results in Obligado \& Vassilicos
\cite{Obligado2019_epl}. The non-stationarity function $f$ is shown in
figure \ref{fig:1b} for $Re_{\lambda}=50,380$. The comparison shows
that there is reasonable agreement for low to moderate Reynolds
numbers, in particular for small values of $r$. The EDQNM prediction
appears to increase slightly faster towards the asymptotic value
$f=4/3$ for large $r$ values. However, the difference between EDQNM
results and grid turbulence data is small, not larger than $\approx
10\%$ in magnitude on this non-stationarity/non-equilibrium
function. This validation supports the view that one might reasonably
expect EDQNM data for much higher Reynolds numbers (up to
$Re_{\lambda} =10^6$) to provide an estimation of the behavior of the
physical quantities investigated, such as $f$ and $F$.

\begin{figure}[h!]
\begin{tabular}{cc}
\includegraphics[width=0.48\linewidth]{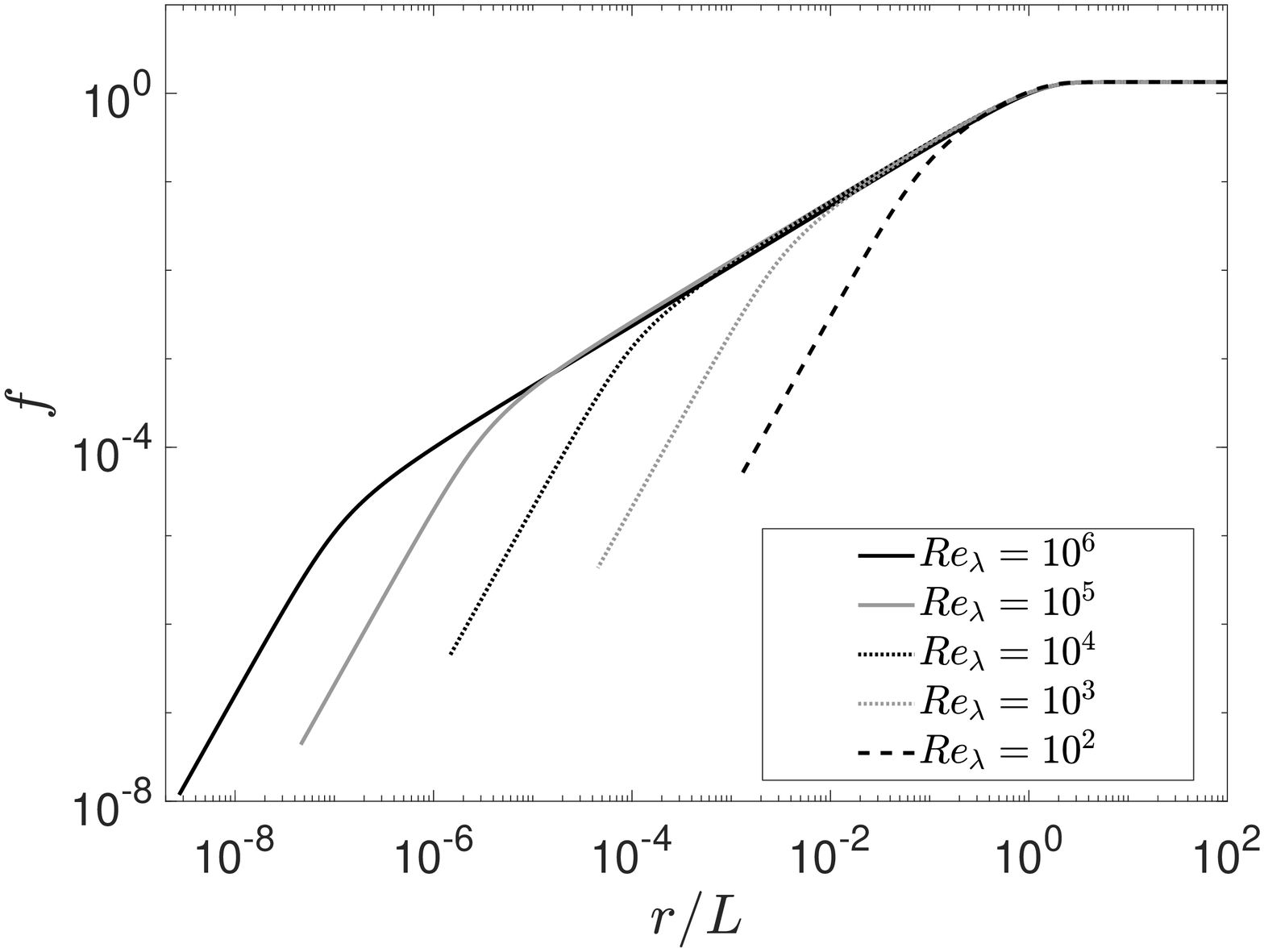} & \includegraphics[width=0.48\linewidth]{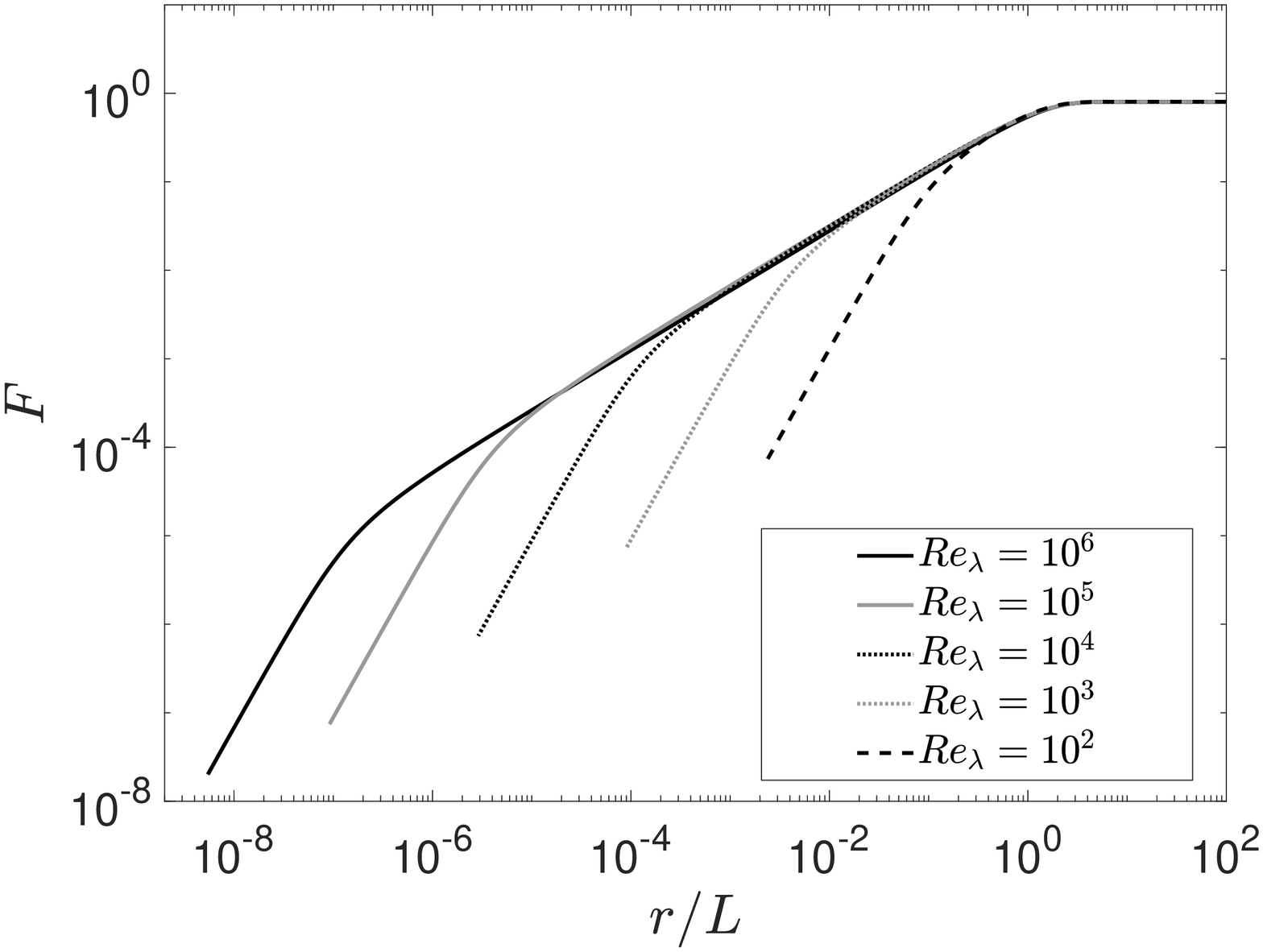} \\
(a) & (b) \\
\includegraphics[width=0.48\linewidth]{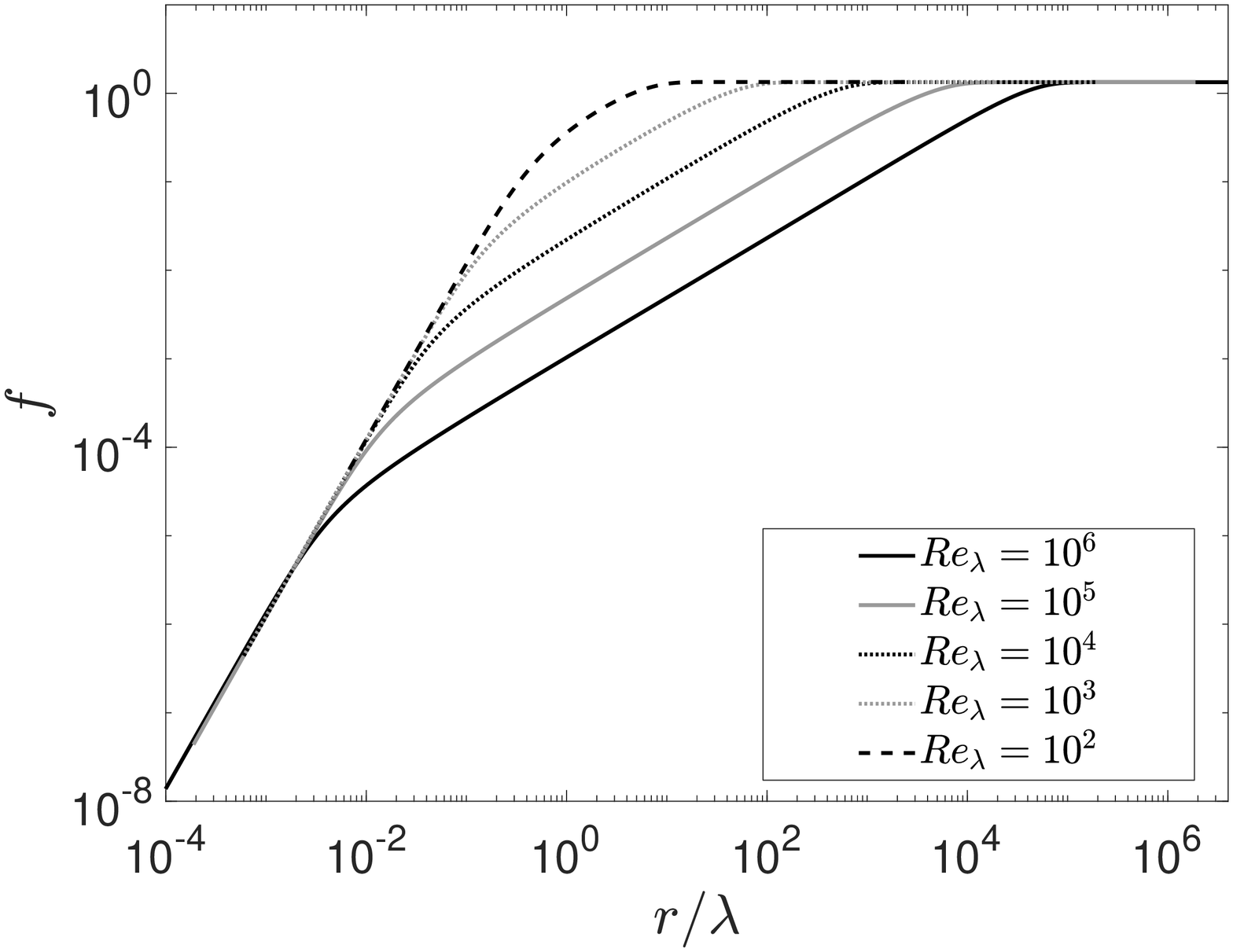} & \includegraphics[width=0.48\linewidth]{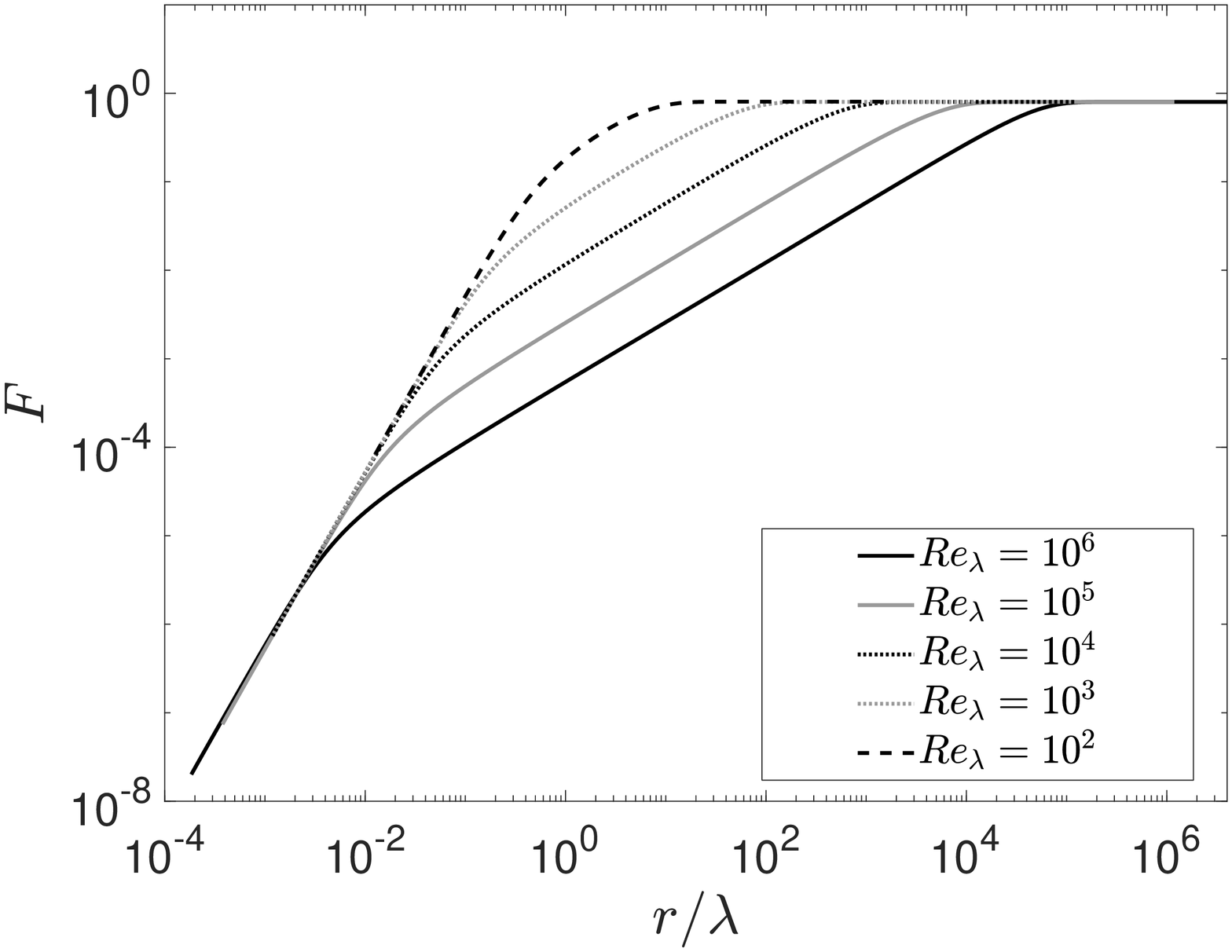} \\
(c) & (d) \\
\includegraphics[width=0.48\linewidth]{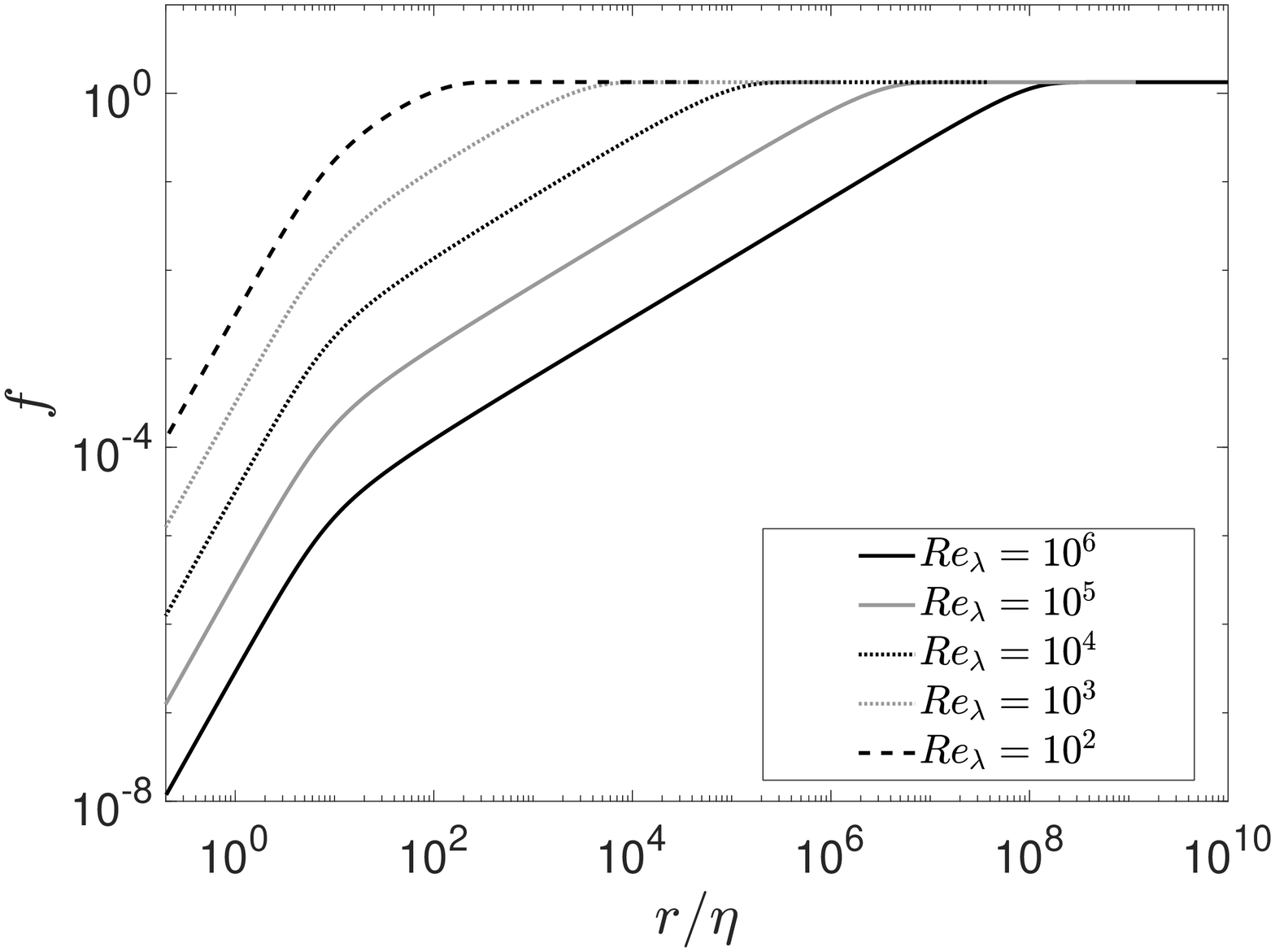} & \includegraphics[width=0.48\linewidth]{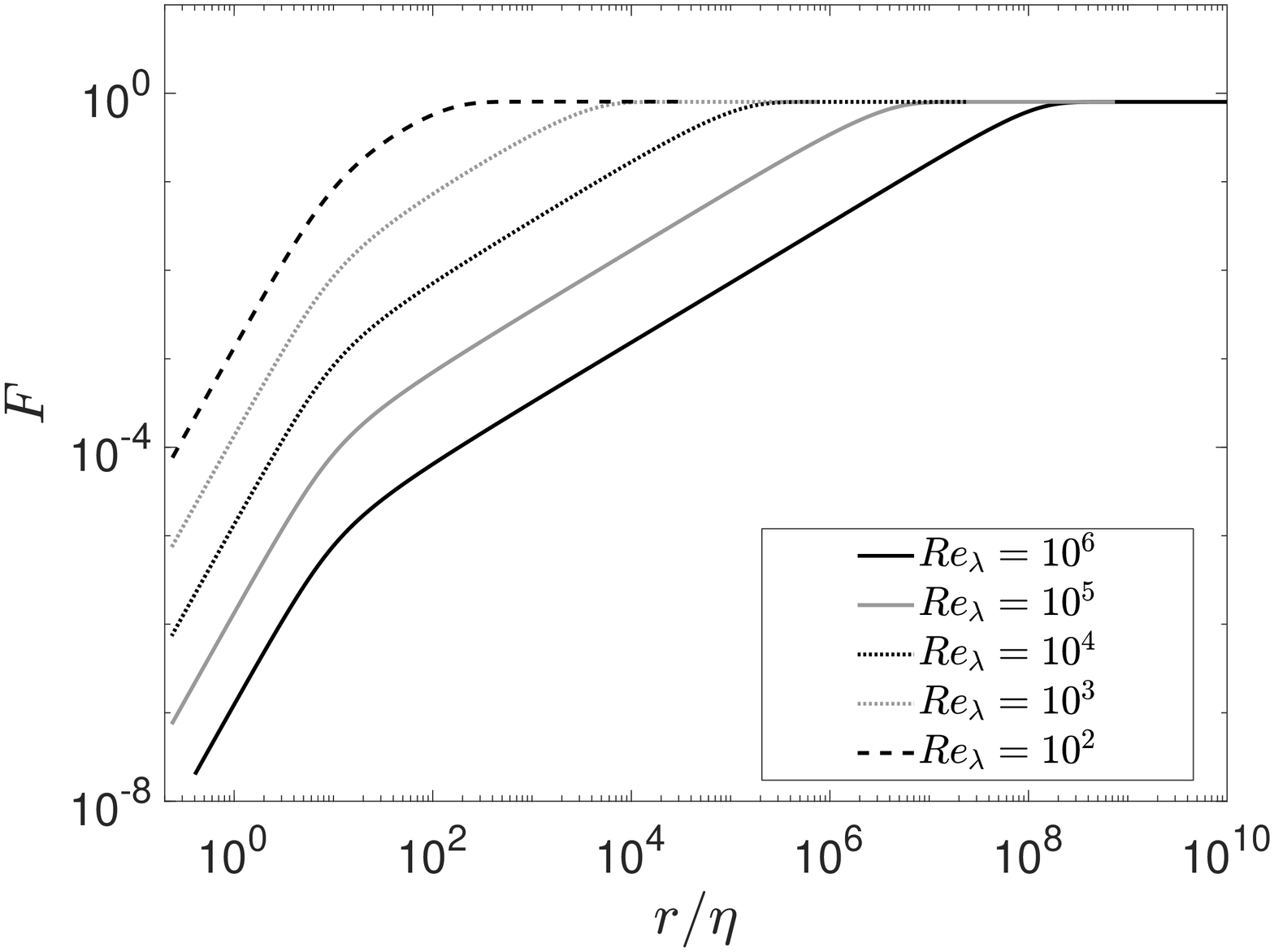} \\
(e) & (f)
\end{tabular}
\caption{\label{fig:2} Non-stationarity functions (left column) $f$ and (right column) $F$ represented against (first row) $r/L$, (second row) $r/ \lambda$ and (third row) $r / \eta$.}
\end{figure}

The EDQNM predictions for $f$ and $F$ are shown in figure
\ref{fig:2}. In the left column, $f$ is plotted against $r/L$,
$r/\lambda$ and $r/\eta$ from top to bottom. Two asymptotic behaviors
are observed for small and large values of $r$. (i) One can see that
$f$ converges towards a constant value of $f=4/3$ for $r > 2 L$ in
agreement with basic theory (see \cite{Obligado2019_epl}). The way
the EDQNM profiles reach the asymptote is the same in terms of $r/L$
as the curves collapse significantly well even before reaching the
asymptotic value in \ref{fig:2}(a). (ii) On the other hand, $f \propto
r^2$ at the small scales which also agrees with basic theory \cite{Taylor1935_prsl}. This range appears to be fully established for $r \le 5 \eta$
approximately (see figure \ref{fig:2}(e)). However, the curves do not
collapse for any value of $r$ when plotted against $r/
\eta$. Self-preservation at the small scales is instead obtained when
using a $r / \lambda$ horizontal axis, as shown in figure
\ref{fig:2}(c). This result confirms that the EDQNM prediction comply
very well with power-law turbulence decay and the theoretical result
$f \approx (r / \lambda)^2 (- \frac{\mathcal{K}}{\varepsilon^2}\frac{d
  \varepsilon}{dt})$ in the small scale region \cite{Taylor1935_prsl,Obligado2019_epl}. The power-law decay implies that $(-
\frac{\mathcal{K}}{\varepsilon^2}\frac{d \varepsilon}{dt})$ is
independent of time and that the entire time-dependence and/or Reynolds
number dependence of $f$ at scales $r \le 5 \eta$ is therefore
captured by $f \approx (r / \lambda)^2$, hence the collapse in figure
\ref{fig:2}(c) which EDQNM is able to reproduce.
 
If the Reynolds number is sufficiently high, a scaling range where $f
\propto r^{2/3}$ emerges between the large scale region and the small
scale region. This range can be qualitatively observed for
$Re_{\lambda} \geq 10^3$. In addition, the scaling predicted by EDQNM
in this range complies very well with the formula $f= -1.5 (\frac{d
  \varepsilon}{dt}/ \varepsilon)(r/ \sqrt{\varepsilon})^{2/3}$ which
follows from $S_{2} \approx {9\over 4} (\varepsilon r)^{2/3}$.

The analysis of the non-stationarity function $F$, which is reported
on the right column of figure \ref{fig:2}, allows to draw almost
identical conclusions. The only observable difference is the value of
the asymptotic limit in the large scale region, which is $F = 4/5$ in
agreement with basic theory \cite{Obligado2019_epl}.

\begin{figure}[h]
\center
\includegraphics[width=0.72\linewidth]{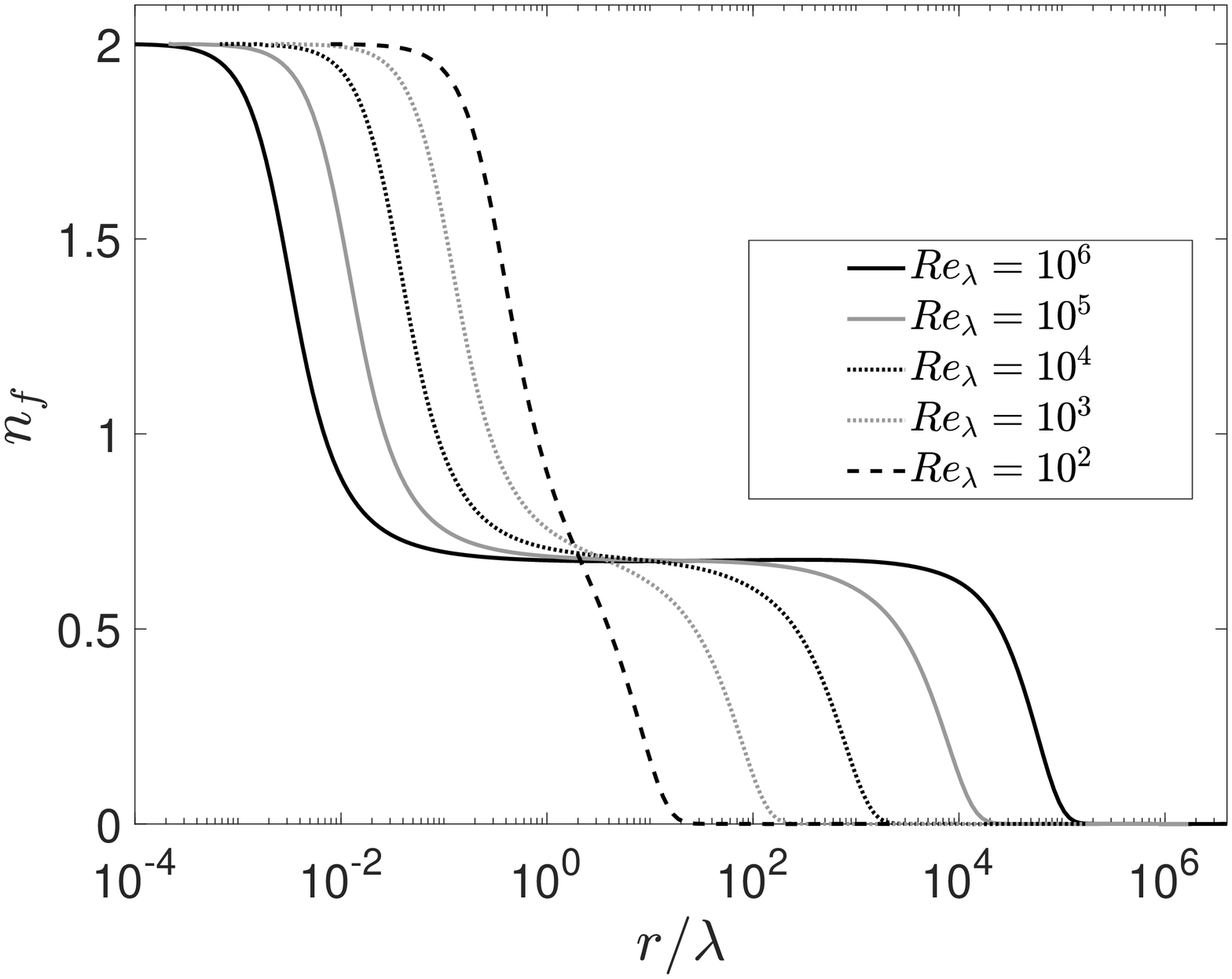} 
\caption{\label{fig:3} Exponent for a power-law $r$-dependence of the
  non-stationarity function $f$. The function is represented as $f
  \propto r^{n_f (r)}$, where $n_f (r)$ is the local power law
  exponent. It is determined via local logarithmic polynomial
  fitting.}
\end{figure}

Results are further analyzed to obtain more accurate information about
the features of the ranges approximately identified. In figure
\ref{fig:3} the exponent $n_f(r)=d \, log(f)/ d \, log(r)$ is
reported: it is calculated via polynomial fitting from the EDQNM
results. This parameter indicates that a scaling range for which $f
\propto r^{2/3}$ is observed for at least one decade only for
$Re_{\lambda} \geq 10^4$. In addition, if one excludes the curve for
the lowest Reynolds number $Re_{\lambda}=10^2$, the inflection point
for $n_f$ is at $r / \lambda \approx 5$ for all Reynolds numbers.

\begin{figure}[h]
\center
\includegraphics[width=0.72\linewidth]{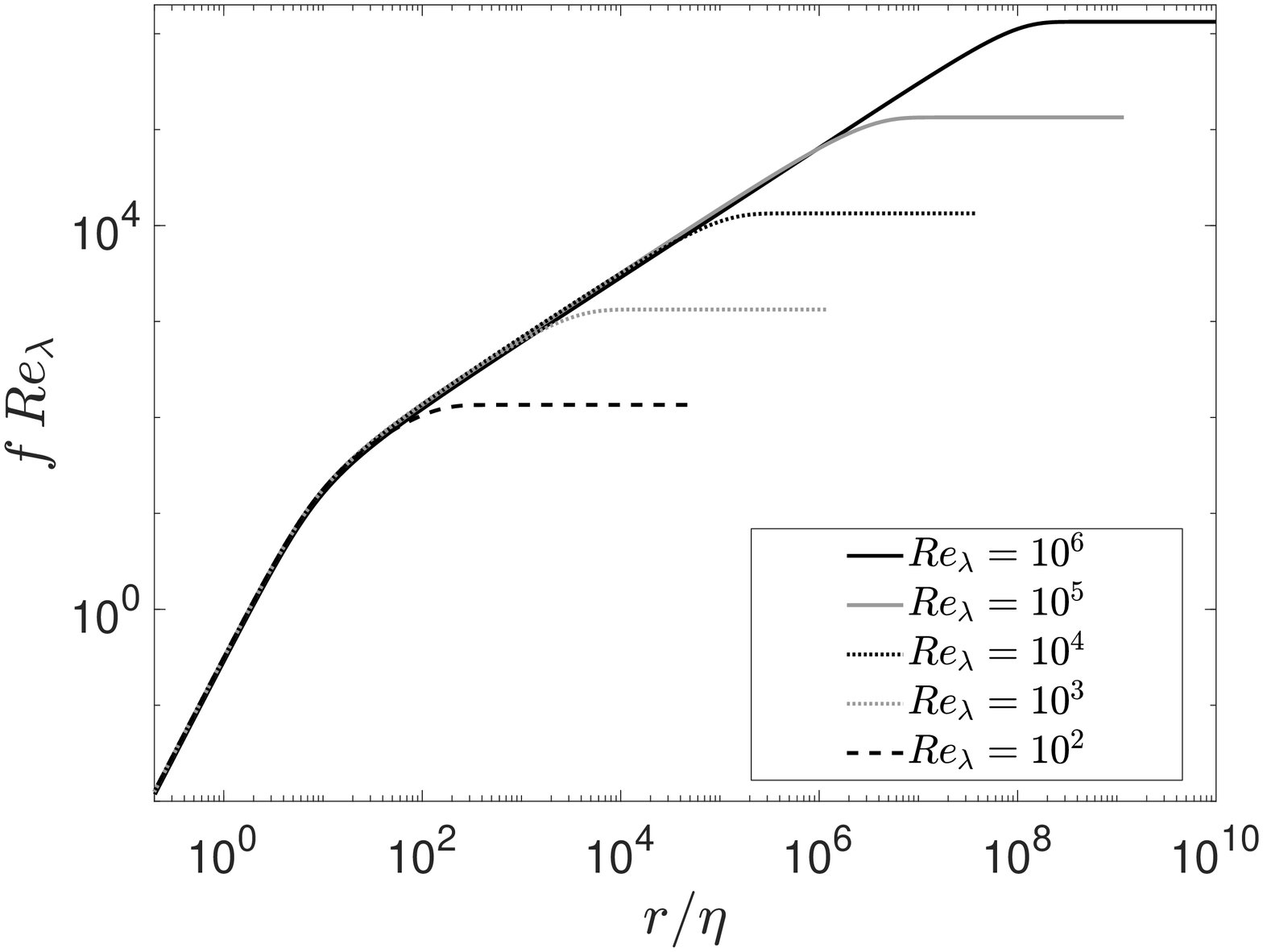} 
\caption{\label{fig:4} Non-stationarity function $f$ times the Reynolds number $Re_{\lambda}$.}
\end{figure}

As already observed, $f$ profiles do not collapse when plotted against
$r/ \eta$ and they exhibit self-preservation in the small scale region
when plotted against $r/\lambda$. However, given that $\lambda / \eta
\propto Re_{\lambda}^{1/2}$, it follows from $f \approx (r /
\lambda)^2 (- \frac{\mathcal{K}}{\varepsilon^2}\frac{d
  \varepsilon}{dt})$ that $f Re_{\lambda} \approx (r / \eta)^2 (-
\frac{\mathcal{K}}{\varepsilon^2}\frac{d \varepsilon}{dt})$ where $(-
\frac{\mathcal{K}}{\varepsilon^2}\frac{d \varepsilon}{dt})$ is
independent of time and Reynolds number for a power-law turbulence
decay. The EDQNM prediction complies with this formula, which is shown
in figure \ref{fig:4}. The function $f Re_{\lambda}$ exhibits partial
self-preservation in the small scale region but also in the
intermediate scaling range when plotted against $r / \eta$.


\begin{figure}[h]
\begin{tabular}{cc}
\includegraphics[width=0.48\linewidth]{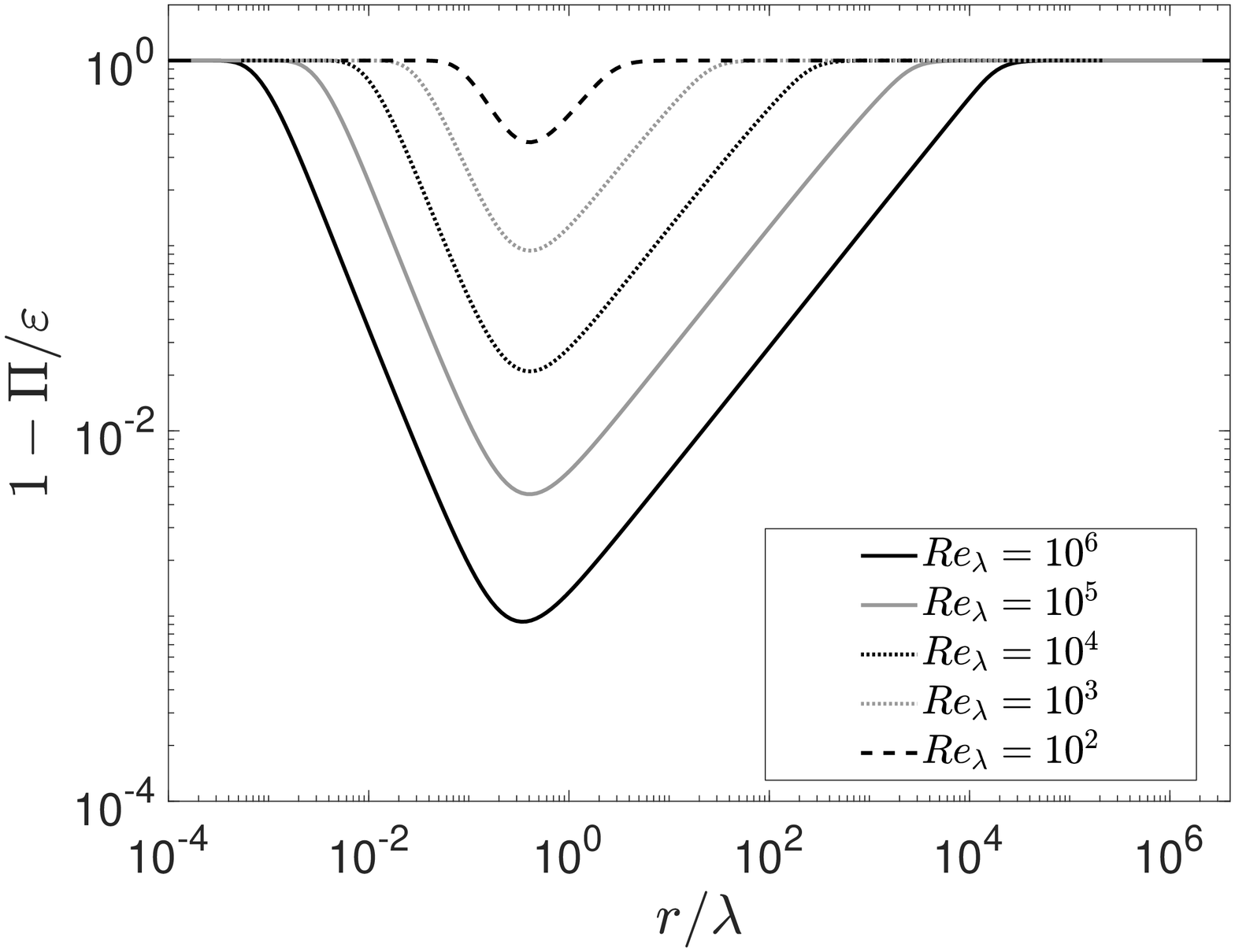} & \includegraphics[width=0.48\linewidth]{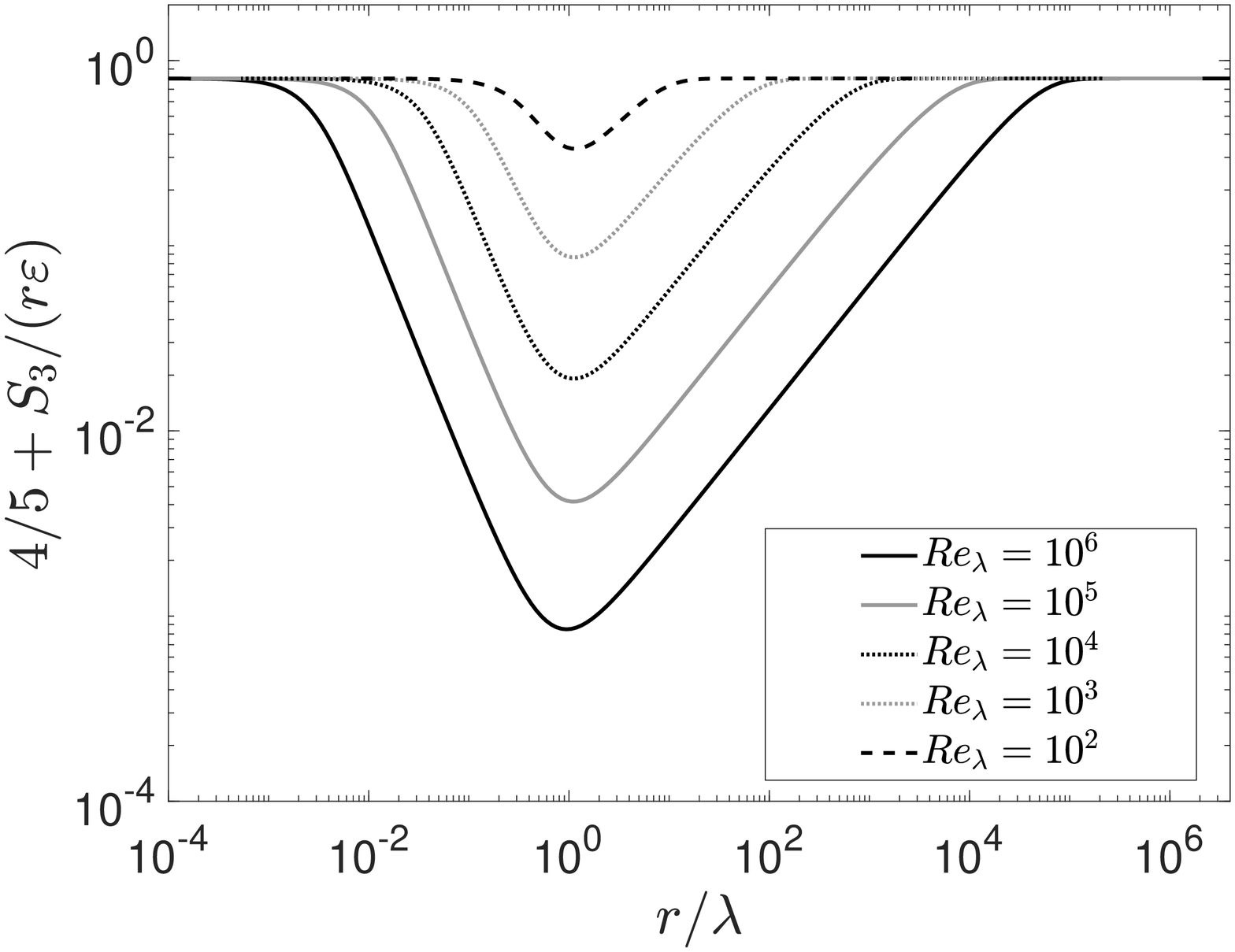} \\
(a) & (b)
\end{tabular}
\caption{\label{fig:5} (a) $1 - \Pi / \varepsilon$ and (b)
  $4/5+S_3/(\varepsilon r)$ represented in log-log scale.}
\end{figure}

\begin{figure}[h]
\begin{tabular}{cc}
\includegraphics[width=0.48\linewidth]{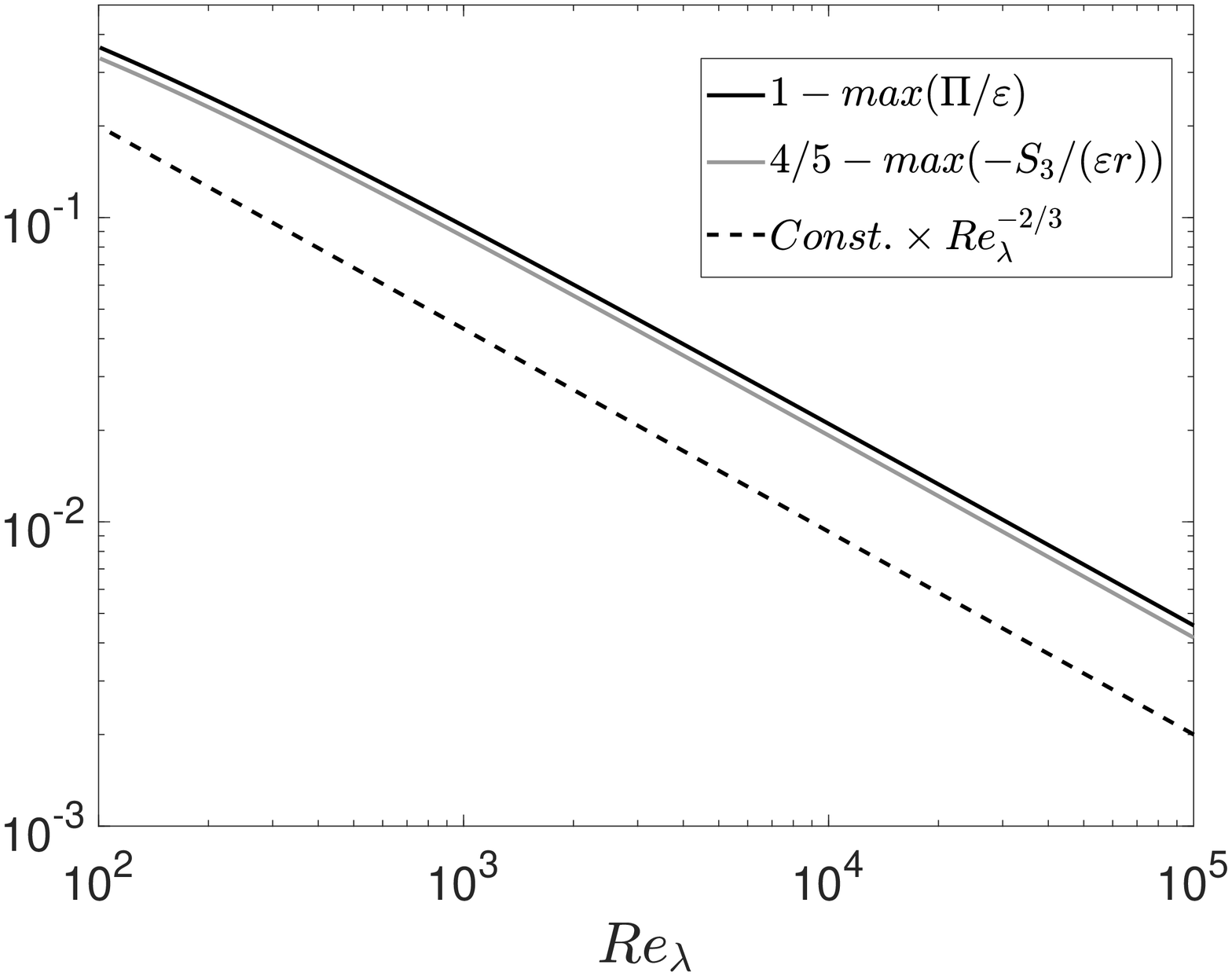} & \includegraphics[width=0.48\linewidth]{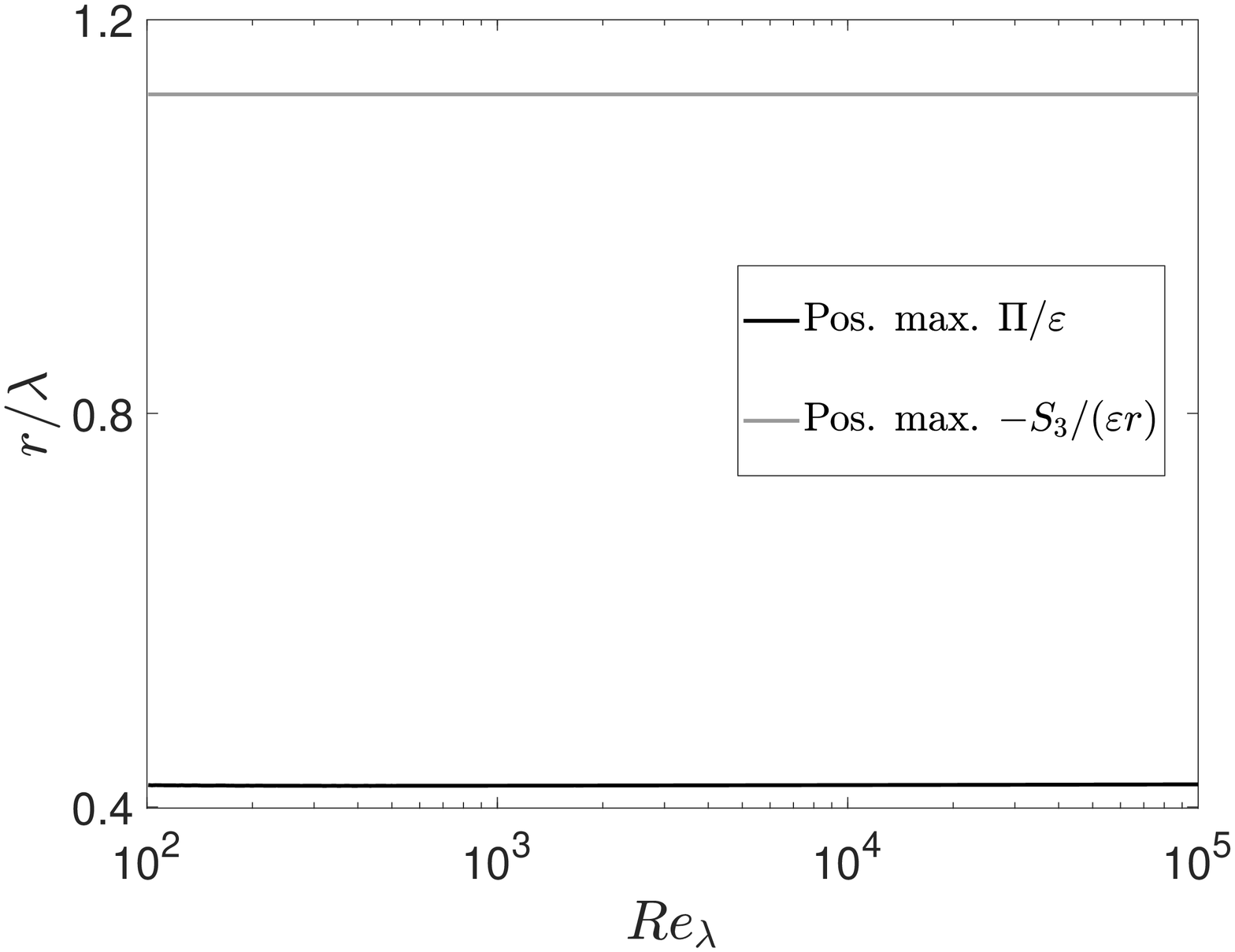} \\
(a) & (b)
\end{tabular}
\caption{\label{fig:6} (a) Calculated values (for each $E$ and $T$
  sampled in $Re_{\lambda} \in [10^2, 10^5]$) of $1- max(\Pi /
  \varepsilon)$ and $4/5-max(-S_3/(\varepsilon r))$. (b) Value of $r /
  \lambda$ where minima are observed.}
\end{figure}

We now investigate detailed EDQNM predictions for the interscale
energy flux $\Pi$ and the third order structure function $S_3$ which
are central quantities for characterising the equilibrium or
non-equilibrium of the turbulence cascade. The function $1-\Pi /
\varepsilon$ is plotted in logarithmic scale in figure
\ref{fig:5}(a). According to Kolmogorov equilibrium, the value of this
function should be zero in some inertial range if the Reynolds number
is large enough. However, the equilibrium $\Pi / \varepsilon = 1$ is
{never observed over an inertial
  (i.e. viscosity-independent) range} for any Reynolds number. The
value of $r$ where $\Pi / \varepsilon$ is closest to 1 turns out to be
$r= 0.42 \lambda$ for every $Re_{\lambda}$ in the range
investigated. Indeed, the minimum for $1 - \Pi / \varepsilon$ is
observed for $r= 0.42 \lambda$ irrespective of Reynolds number in
figure \ref{fig:6}(a). As $Re_{\lambda}$ increases towards infinity,
$1 - \Pi / \varepsilon$ tends to $0$ only for values of $r$ which
remain a fixed factor or multiple of $\lambda$ as $Re_{\lambda}$
varies. In other words the tendency towards equilibrium is only
present in the vicinity of the Taylor scale $\lambda$ in perfect
agreement with the prediction of Lundgren \cite{lundgren2002} and
Obligado \& Vassilicos \cite{Obligado2019_epl}. As the Taylor scale
depends on viscosity we cannot say that the tendency towards
equilibrium happens in some inertial range. As figure \ref{fig:5}(a)
suggests, in a range of scales which are independent of viscosity the
degree of non-equilibrium persists with increasing Reynolds number and
the closer the length-scale $r$ is to $L$ the further the turbulence
is from equilibrium. Results in figure \ref{fig:5}(a) indicate that
this departure from equilibrium scales as $(r/\lambda)^{2/3}$ for $r>
\lambda$ and as $r/\lambda$ for
$r<\lambda$. {These formulae can be used to
  determine the span of an empirical range respecting the condition
  $\Pi / \varepsilon \approx 1$, down to a prescribed tollerance. In
  fact, if one fixes a level of tollerance $L_T$ tracing a horizontal
  line in figure \ref{fig:5}(a) (let's say for example $L_T=10^{-2}$,
  which corresponds to an error of $1\%$), an approximated triangle in
  the spectral space can be drawn intersecting this line with the
  curve for a selected $Re_{\lambda}$. Using the scalings introduced
  two sentences above this one, and considering that the vertical
  position of the lower vertex of the triangle can be approximated by
  the value $min(1-\Pi / \varepsilon)$, one can determine the
  extension of this empirical range $\Delta_{er}$:}
\begin{equation}
\Delta_{er}= 10^{\sigma_{er}} \, , \qquad \sigma_{er} = \left(\log_{10}(P_r) - \log_{10}(min(1-\Pi / \varepsilon)) \right) \left(1 + \frac{3}{2} \right)
\label{eq:empiricalRange}
\end{equation}
{The coefficient (1+3/2) on the right is obtained
  from the observation of slopes $r/\lambda$ for $r<\lambda$ and
  $(r/\lambda)^{2/3}$ for $r>\lambda$ in figure \ref{fig:5}(a). The
  span of this range increases with $Re_{\lambda}$ but it also
  significantly changes with the degree of tollerance imposed for the
  relation $\Pi / \varepsilon \approx 1$. For high $Re_{\lambda}$ and
  relatively low tollerance, one can obtain a significantly large
  range which will not be exactly centered around $\lambda$, as the
  function $1 - \Pi / \varepsilon$ exhibits different evolutions
  moving from $\lambda$ towards the small scales and the large scales
  respectively. Such a coarse analysis may erroneously lead one to
  think that the condition $\Pi / \varepsilon \approx 1$ holds in some bit of an inertial range, but the fact remains that $\Pi / \varepsilon$ approaches $1$ as $Re_{\lambda} \to \infty$ only for scales $r$ around $\lambda$, i.e. in a range of scales which is not inertial.}

{In conclusion, the EDQNM model} agrees with the
matched asymptotic expansion analysis of the K\'arm\'an-Howarth
equation of Lundgren \cite{lundgren2002} (confirmed by the wind tunnel
data analysis of Obligado \& Vassilicos \cite{Obligado2019_epl}) which
predicts that there is no inertial range with an approximate
equilibrium between $\Pi$ and $\varepsilon$ and that, instead, there
is a systematic departure from equilibrium as scales $r$ are
considered further and further away from
$\lambda$. {For any fixed $r/L$ value, however
  small, $\Pi / \varepsilon$ does not tend to 1 as $Re_{\lambda}\to
  \infty$, even if $r/L$ is so very small that $\Pi / \varepsilon$
  is close to 1.}


Similar conclusions can be drawn for $S_3$, which is shown in figure
\ref{fig:5}(b). Here the Kolmogorov equilibrium value $S_3/(
\varepsilon r)=-4/5$ for HIT is never reached, except asymptotically
as $Re_{\lambda} \to +\infty$ at a value of $r/\lambda$ which remains
fixed as $Re_{\lambda}$ grows. In fact, Lundgren's (2002) matched
asymptotic expansion analysis of the K\'arm\'an-Howarth equation
predicts that the minimum value of $4/5 + S_3 / (\varepsilon r)$ is
reached at a value of $r$ which is a fixed multiple of $\lambda$ and
that $4/5 -max(-S_3 / (\varepsilon r)) \propto Re_{\lambda}^{-2/3}$
(see Obligado \& Vassilicos 2019). Our EDQNM results are in full
agreement with these predictions and return a value $r=1.12\lambda$
(see figure \ref{fig:6}(b)) for the scale $r$ where $4/5 + S_3 /
(\varepsilon r)$ is minimal and a decay towards 0 of this minimal
value which is indeed proportional to $Re_{\lambda}^{-2/3}$ as
$Re_{\lambda}$ increases (see figure
\ref{fig:6}(a)). {More precisely, EDQNM results
  suggest that the minimum can be well approximated by the relation
  $min(1-\Pi / \varepsilon) \approx 10 Re_{\lambda}^{-2/3}$.}

EDQNM results for the function $1-max(\Pi/\varepsilon)$ are also shown
in figure \ref{fig:6}(a) in the range $Re_{\lambda} \in [10^2, \,
  10^5]$. Reflecting the behaviour of $S_3$, $1-max(\Pi/\varepsilon)
\propto Re_{\lambda}^{-2/3}$ as $Re_{\lambda} \to +\infty$ and the
value of $r$ where $1-(\Pi/\varepsilon)$ takes its minimum value also
scales with $\lambda$: figure \ref{fig:6}(b) shows that this value of
$r$ is $r= 0.42 \lambda$ for every $Re_{\lambda}$ in the range
investigated.

\section{Sensitivity analysis}
\label{sec:sensitivity}

{The sensitivity of the results shown in figure
  \ref{fig:6} is now analyzed by comparing results from different
  simulations of our database in order to quantify the effects of the
  variations in the parametric description.}

\begin{figure}[h]
\begin{tabular}{cc}
\includegraphics[width=0.48\linewidth]{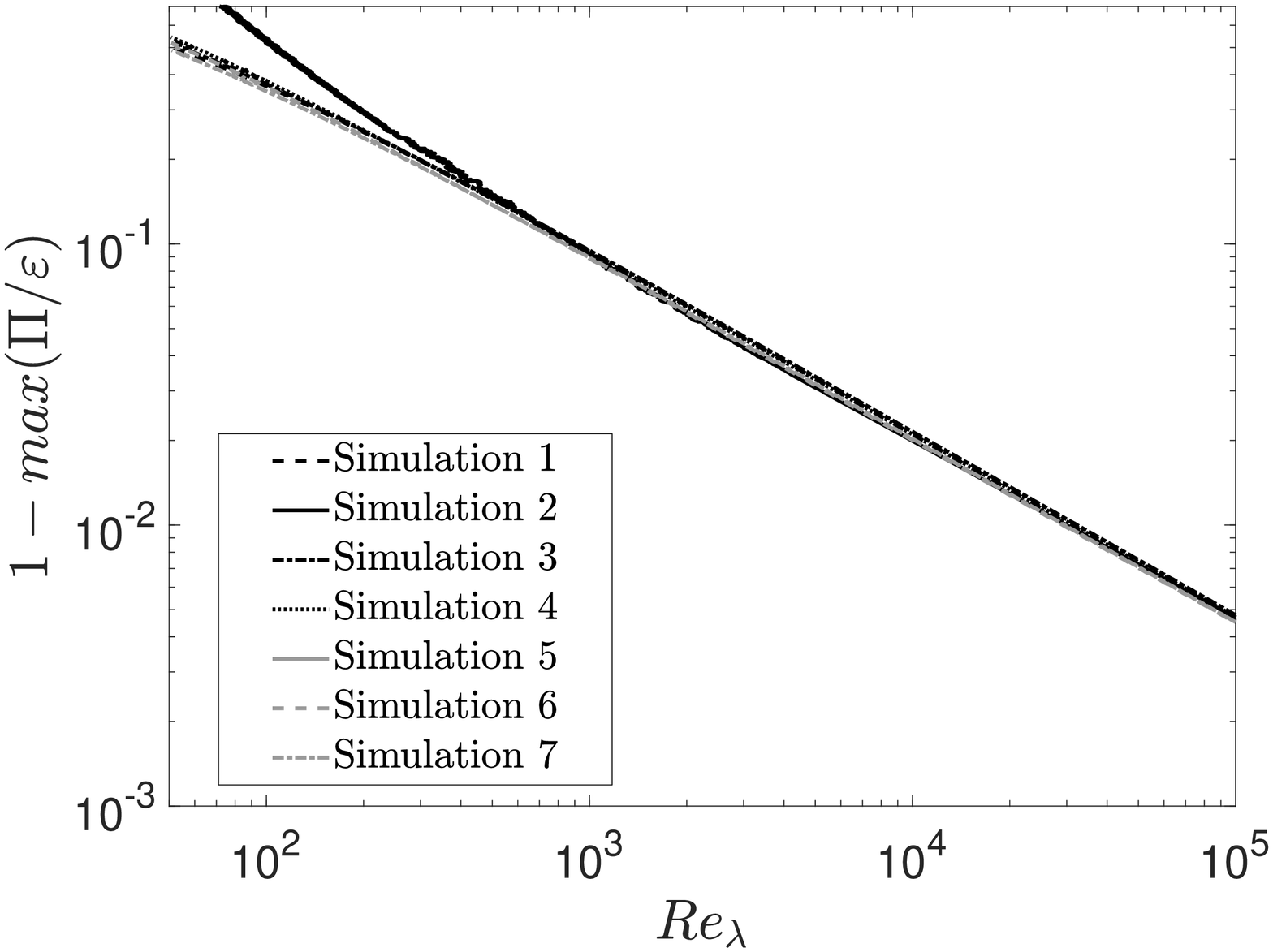} & \includegraphics[width=0.48\linewidth]{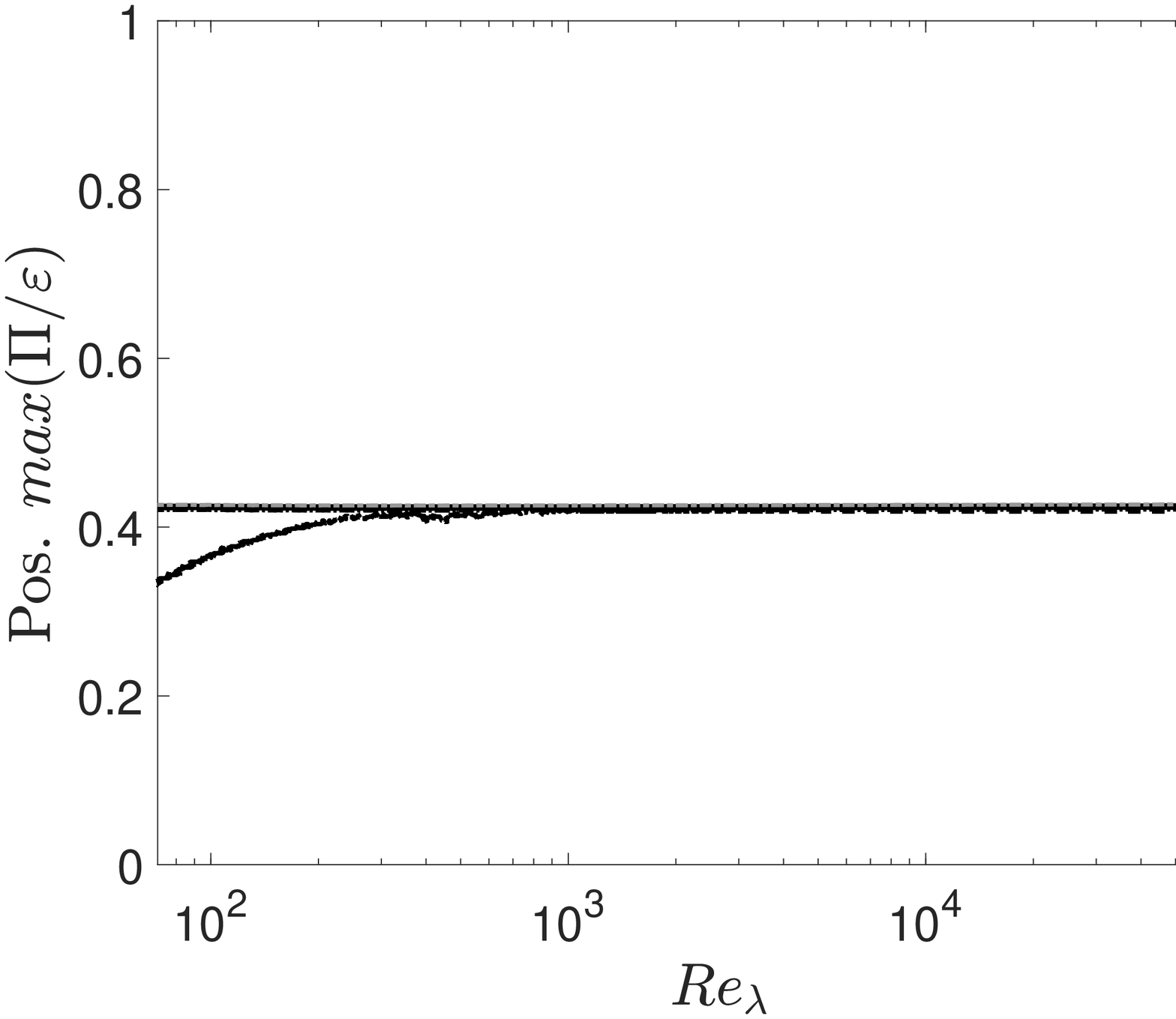} \\
(a) & (b)
\end{tabular}
\caption{\label{fig:7} {Sensitivity of (a) calculated values of $1- max(\Pi /
  \varepsilon)$ and (b) value of $r / \lambda$ where minima are observed, to the large scale features imposed in the EDQNM calculations.}}
\end{figure}

{More specifically, the sensitivity on the
  parameters characterising the large scale features is investigated
  for $min(1-\Pi / \varepsilon)$ and for the value of $r / \lambda$
  where minima are observed. This sensitivity analysis is based on
  results from the simulations $1$ to $7$ in our database (see table
  \ref{table::1} for more information). The large-scale features
  encompass the actual parameters governing the large scales, such as
  $\sigma$, but also the two EDQNM models and the functional forms
  prescribed for $E(k,t=0)$. The results, which are shown in figure
  \ref{fig:7}, show that variations in $\sigma$, in the prescribed
  form for $E(k,t=0)$ and in the model for the eddy-damping term in
  EDQNM do not significantly affect the results. However, some
  differences may be observed for simulation 2 at $Re_{\lambda} <
  400$. For this simulation, confinement effects at the large scales
  are more important than in the other calculations in the
  database. As $Re_{\lambda}$ decreases, the reduced scale separation
  allows the confinement effects to affect $min(1-\Pi / \varepsilon)$
  and the value of $r / \lambda$ where this minimum is observed.}

\section{Conclusions}
\label{sec:conclusions}

It is perhaps remarkable how well EDQNM of freely decaying HIT agrees
with the results of Lundgren's matched asymptotic expansion approach
\cite{lundgren2002} to the K\'arm\'an-Howarth equation. Both predict
that a Kolmogorov inertial range equilibrium is not
obtained. Equilibrium is only achieved in the vicinity of the Taylor
length in the limit where the Reynolds number tends to infinity and
the Taylor length is not in the inertial range given its explicit
dependence on viscosity. As the length-scale $r$ moves away from
$\lambda$ and towards the integral scale $L$, the turbulence moves
progressively further and further away from Kolmogorov equilibrium.
The inertial range is in fact characterised by a gradually increasing
departure from equilibrium as $r$ increases from $\lambda$ to $L$,
{not by uniform near-equilibrium over much or
  even a significant part of this range}, however large the Reynolds
number may be. {At a length-scale $r$ taken to be
  a fixed small fraction (however small) of the integral scale, the
  ratio of the interscale energy flux to the turbulence dissipation
  does not tend to 1 a the Reynolds number tends to infinity.}

Similar considerations can be extended to the case where $r$ decreases
from $\lambda$ to $\eta$ {except that there is no
  inertial range at these very small length-scales}. However, this
last range is significantly shorter and it can be clearly observed
only for very high $Re_{\lambda}$.


This raises the question of justifying the presence of the
Taylor-Kolmogorov scaling of the turbulence dissipation even though
there is no equilibrium in the inertial range, particularly at the
larger-scale end of it. Goto \& Vassilicos \cite{Goto2016_pre}
introduced the concept of balanced non-equilibrium for such a
justification. They considered the Lin equation integrated over
wavenumbers from $k$ to infinity and defined balanced non-equilibrium
to mean that the 3 terms in this integrated Lin equation vary together
in time for wavenumbers $k \sim 1/L$. Balanced non-equilibrium
therefore means that
\begin{equation}
\label{eq:bala}
    \frac{\partial }{\partial t}\int_{1/L}^{+\infty}E(k,t)dk \sim 2
    \nu \int_{1/L}^{+\infty}k^2 E(k,t) dk \sim \Pi(1/L,t)
\end{equation}
which leads to the Taylor-Kolmogorov dissipation scaling given that $2
\nu \int_{1/L}^{+\infty}k^2 E(k,t) dk \approx \varepsilon$ and that
$\Pi(1/L,t)$ can be considered to be independent of inlet/initial
conditions and viscosity. The validity of balanced non-equilibrium was
demonstrated in the context of EDQNM by Meldi \& Sagaut
\cite{Meldi2013_jot} and therefore explains the fact that EDQNM
returns the Taylor-Kolmogorov scaling for the turbulence dissipation
as reported in previous publications
\cite{Bos2007_pof,Meldi2013_jot,Sagaut2018_springer}. {Future potential investigations could employ more sophisticated tools able to take into account anisotropy of the flow, such as specific versions of the EDQNM model reported in the literature \cite{Mons2014_pof,Mons2015_jfm,Briard2016_jfm}.}

\bibliographystyle{unsrt} \bibliography{references1}

\begin{thebibliography}{10}

\bibitem{Vassilicos2015_arfm}
J.C. Vassilicos.
\newblock {Dissipation in turbulent flows}.
\newblock {\em Annual Review of Fluid Mechanics}, 47:95--114, 2015.

\bibitem{Cafiero2019_prsa}
G.~Cafiero and J.~C. Vassilicos.
\newblock {Non-equilibrium turbulence scalings and self-similarity in turbulent
  planar jets}.
\newblock {\em Proceedings of the Royal Society A}, 475 (2225):20190038, 2019.

\bibitem{Taylor1935_prsl}
G.~I. Taylor.
\newblock {Statistical Theory of Turbulence}.
\newblock {\em Proceedings of the Royal Society of London A}, 151:421--444,
  1935.

\bibitem{Kolmogorov1941_dan}
A.~N. Kolmogorov.
\newblock {On the degeneration of isotropic turbulence in an incompressible
  viscous fluid}.
\newblock {\em Doklady Akademii nauk SSSR}, 31(6):538 -- 541, 1941.

\bibitem{Antonia2006}
R.A. Antonia and P.~Burattini.
\newblock Approach to the 4/5 law in homogeneous isotropic turbulence.
\newblock {\em J.~Fluid Mech.}, 550:175--184, 2006.

\bibitem{Goto2016_pre}
S.~Goto and J.~C. Vassilicos.
\newblock {Unsteady turbulence cascades}.
\newblock {\em Physical Review E}, 94:053108, 2016.

\bibitem{Obligado2019_epl}
M.~Obligado and J.~C. Vassilicos.
\newblock {The non-equilibrium part of the inertial range in decaying
  homogeneous turbulence}.
\newblock {\em Europhysics Letters}, 127:64004, 2019.

\bibitem{Karman1938_prsA}
T.~Von~Karman and L.~Howarth.
\newblock {On the statistical theory of isotropic turbulence}.
\newblock {\em Proceedings of the Royal Society A}, 164:192--215, 1938.

\bibitem{Landau1987}
L.~D. Landau and E.~M. Lifshitz.
\newblock {\em Fluid mechanics: Vol 6 of course of theoretical physics, Second
  Edition}.
\newblock Pergamon Press, 1987.

\bibitem{lundgren2002}
Thomas~S Lundgren.
\newblock Kolmogorov two-thirds law by matched asymptotic expansion.
\newblock {\em Phys.~ Fluids}, 14(2):638--642, 2002.

\bibitem{Orszag1970_jfm}
S.~A. Orszag.
\newblock {Analytical theories of turbulence}.
\newblock {\em J.~Fluid Mech.}, 41:363 -- 386, 1970.

\bibitem{Bos2007_pof}
W.~J.~T. Bos, L.~Shao, and J.~P. Bertoglio.
\newblock {Spectral imbalance and the normalized dissipation rate of
  turbulence}.
\newblock {\em Phys. Fluids}, 19(4):045101, 2007.

\bibitem{Lesieur2008_springer}
M.~Lesieur.
\newblock {\em {Turbulence in Fluids (4th edition)}}.
\newblock Springer, 2008.

\bibitem{Meldi2013_jot}
M.~Meldi and P.~Sagaut.
\newblock Further insights into self-similarity and self-preservation in freely
  decaying isotropic turbulence.
\newblock {\em Journal of Turbulence}, 14:24--53, 2013.

\bibitem{Sagaut2018_springer}
P.~Sagaut and C.~Cambon.
\newblock {\em {Homogenous Turbulence Dynamics}}.
\newblock Springer Verlag, 2018.

\bibitem{Pouquet1975_jfm}
A.~Pouquet, M.~Lesieur, J.-C. Andr\'e, and C.~Basdevant.
\newblock Evolution of high reynolds number two-dimensional turbulence.
\newblock {\em J.~ Fluid Mech.}, 75:305--319, 1975.

\bibitem{Andre1977_jfm}
J.-C. Andr\'e and M.~Lesieur.
\newblock Influence of helicity on the evolution of isotropic turbulence at
  high reynolds number.
\newblock {\em J.~ Fluid Mech.}, 81:187--207, 1977.

\bibitem{Bos2006_pof}
W.~J.~T. Bos and J.-P. Bertoglio.
\newblock A single-time two-point closure based on fluid particle
  displacements.
\newblock {\em Phys.~ Fluids}, 18:031706, 2006.

\bibitem{Meldi2014_jcp}
M.~Meldi and P.~Sagaut.
\newblock {An adaptive numerical method for solving EDQNM equations for the
  analysis of long-time decay of isotropic turbulence}.
\newblock {\em Journal of Computational Physics}, 262:72--85, 2014.

\bibitem{Pope2000_cambridge}
S.~B. Pope.
\newblock {\em {Turbulent Flows}}.
\newblock Cambridge University Press, 2000.

\bibitem{Meyers2008_pof}
J.~Meyers and C.~Meneveau.
\newblock {A functional form for the energy spectrum parametrizing bottleneck
  and intermittency effects}.
\newblock {\em Phys. Fluids}, 20(6):065109, 2008.

\bibitem{Meldi2012_jfm}
M.~Meldi and P.~Sagaut.
\newblock {On non-self-similar regimes in homogeneous isotropic turbulence
  decay}.
\newblock {\em J.~Fluid Mech.}, 711:364--393, 2012.

\bibitem{Meldi2017_jfm}
M.~Meldi and P.~Sagaut.
\newblock {Turbulence in a box: Quantification of large-scale resolution
  effects in isotropic turbulence free decay}.
\newblock {\em J.~Fluid Mech.}, 818:697--715, 2017.

\bibitem{Comte-Bellot1966_jfm}
G.~Comte-Bellot and S.~Corrsin.
\newblock {The use of a contraction to improve the isotropy of grid-generated
  turbulence}.
\newblock {\em J.~Fluid Mech.}, 25:657 -- 682, 1966.

\bibitem{Kolmogorov41a}
A.~N. Kolmogorov.
\newblock The local structure of turbulence in incompressible viscous fluid for
  very large \uppercase{R}eynolds number.
\newblock {\em Dokl. Akad. Nauk SSSR}, 30 (see also Proc. R. Soc. Lond. A
  (1991), 434, 9-13), 1941a.

\bibitem{Kolmogorov41b}
A.~N. Kolmogorov.
\newblock Dissipation of energy in the locally isotropic turbulence.
\newblock {\em Dokl. Akad. Nauk SSSR}, 32 (see also Proc. R. Soc. Lond. A
  (1991), 434, 15-17), 1941b.

\bibitem{Tchoufag2012_pof}
J.~Tchoufag, P.~Sagaut, and C.~Cambon.
\newblock {Spectral approach to finite \uppercase{R}eynolds number effects on
  \uppercase{K}olmogorov's 4/5 law in isotropic turbulence.}
\newblock {\em Phys. Fluids}, 24(1):015107, 2012.

\bibitem{Bos2012_pof}
W.~J.~T. Bos, L.~Chevillard, J.~F. Scott, and R.~Rubinstein.
\newblock Reynolds number effect on the velocity increment skewness in
  isotropic turbulence.
\newblock {\em Phys.~ Fluids}, 24:015108, 2012.

\bibitem{Danaila99}
L.~Danaila, F.~Anselmet, T.~Zhou, and R.~A. Antonia.
\newblock A generalization of \uppercase{Y}aglom's equation which accounts for
  the large-scale forcing in heated decaying turbulence.
\newblock {\em J. Fluid Mech.}, 391:359--372, 1999.

\bibitem{Mons2014_pof}
V.~Mons, M.~Meldi, and P.~Sagaut.
\newblock {Numerical investigation on the partial return to isotropy of freely
  decaying homogeneous axisymmetric turbulence}.
\newblock {\em Phys. Fluids}, 26:025110, 2014.

\bibitem{Mons2015_jfm}
V.~Mons, C.~Cambon, and P.~Sagaut.
\newblock A spectral model for homogeneous shear-driven anisotropic turbulence
  in terms of spherically averaged descriptors.
\newblock {\em J.~ Fluid Mech.}, 788:147--182, 2015.

\bibitem{Briard2016_jfm}
A.~Briard, T.~Gomez, and C.~Cambon.
\newblock Spectral modelling for passive scalar dynamics in homogeneous
  anisotropic turbulence.
\newblock {\em J.~ Fluid Mech.}, 799:159--199, 2016.

\end{thebibliography}

\end{document}